\documentclass[11pt]{article}
\usepackage[utf8]{inputenc}

\usepackage[margin=1in]{geometry}
\newgeometry{vmargin={2cm}, hmargin={12mm,17mm}}
\usepackage[width=0.9\textwidth,font=footnotesize,labelfont=bf,]{caption}
\usepackage{amsmath}
\usepackage{amssymb}
\usepackage{todonotes}
\usepackage{subcaption}
\usepackage[capitalise]{cleveref}
\usepackage[numbers, sort&compress]{natbib}
\usepackage{siunitx}

\usepackage{soul}

\newenvironment{invisibleCases}
{\left.\begin{aligned}}
	{\end{aligned}\right.}
	
\newcommand{\pe}{Pe}
\newcommand{\pecrit}{\pe^\ast_\ell}

\newcommand{\taudk}{\tau_D k_\text{off}}
\newcommand{\cpfc}{c_0 \partial_c f(c_0)}

\DeclareMathOperator*{\argmax}{arg\,max}

\title{A computational model of self-organized shape dynamics of active surfaces in fluids}
\author{Lucas D. Wittwer$^{1,2}$, Sebastian Aland$^{1,2}$}

\date{
$^1$Institut für Numerische Mathematik und Optimierung, TU Freiberg, 09599 Freiberg, Germany\\
$^2$Fakultät Informatik/Mathematik, HTW Dresden, 01069 Dresden, Germany\\
\today
}

\begin{document}

\maketitle
\abstract{
Mechanochemical processes on surfaces such as the cellular cortex or epithelial sheets, play a key role in determining patterns and shape changes of biological systems. 
To understand the complex  interplay of hydrodynamics and material flows on such active surfaces requires novel numerical tools. 
Here, we present a finite-element method for an active deformable surface interacting with the surrounding fluids. 
The underlying model couples surface and bulk hydrodynamics to surface flow of a diffusible species which generates active contractile forces. 
The method is validated with previous results based on linear stability analysis and shows almost perfect agreement regarding predicted patterning. 
Away from the linear regime we find rich non-linear behavior, such as the presence of multiple stationary states. 
We study the formation of a contractile ring on the surface and the corresponding shape changes. 
Finally, we explore mechanochemical pattern formation on various surface geometries and find that patterning strongly adapts to local surface curvature.
The developed method provides a basis to analyze a variety of systems that involve mechanochemical pattern formation on active surfaces interacting with surrounding fluids. 
}

\section{Introduction}
The theory of active matter captures the ability of living systems to generate spontaneous motion, flows, and material deformations as a result of molecular force-generating processes \cite{Marchetti2013-kq, Julicher2018-mn}. 
At the cellular level, such processes are notably present in the actomyosin cortex, a thin polymeric layer connected to the cell membrane. 
Motor proteins, such as myosin, interact with the cortical filaments, thereby generating movement and forces.
This makes the cell surface essentially an active material, governing cell mechanics and cell shape changes \cite{Mayer2010-ex, Salbreux2012-eu}. 
A striking example is found in cell division, during which a ring of high myosin concentrations is formed and constricts the cell to create two daughter cells \cite{Mayer2010-ex,Pollard2010-ft}. Also, the morphogenesis of tissues does involve an interplay between chemical signals and active mechanics of their surrounding surfaces layer \cite{Heisenberg2013-sh, Martin2010-kl}.

To induce the concentration gradients which are necessary to drive any complex motion, spontaneous pattern formation in active surfaces emerges by a generic mechanochemical organization of the motor-proteins \cite{Bois2011-ev}: As their concentration regulates the amplitude of active stress, it induces material flows transporting more motor-proteins to regions of high concentration. This positive feedback loop is further influenced by corresponding changes of the underlying geometry of the surface and surrounding media. 
To fundamentally understand how active stresses are dynamically organized during shape changes of cells and tissues requires advanced physical and numerical modeling. 

Self-organized active fluids have been studied previously on fixed domains \cite{Bois2011-ev, Kumar2014-gw, Moore2014-dm, Sehring2015-hg, Weber2018-nn, Mietke2019-qf}. To our knowledge, the only method which considers deforming active surfaces was developed in \cite{Mietke2019-ju}. This method couples surface hydrodynamics to material flow and axisymmetric surface deformations. It was found that the system gives rise to polarization with single concentration spots. Ring-like patterns which could explain cellular constriction during cell division were found to be suppressed by cellular polarization. Hence, to obtain ring patterns around the surface requires taking additional components of the biophysical system into account. One of which is the rheology of the surrounding material, which can give rise to more complex patterns, as demonstrated by Mietke at al. \cite{Mietke2019-qf}. Including the viscous interior of the cell, they found stable ring-like patterns by linear stability analysis and numerical simulations. But as the numerical and analytical model are designed for stationary spherical surfaces only, the interaction between self-organization and shape changes remains elusive. 

In this paper, we develop the first numerical method which combines hydrodynamics of a deformable active surface and the surrounding viscous media. The model is based on a general force balance of a curved fluidic surface embedded in viscous fluids, as well as constitutive relations that describe the active surface force in dependence on the surface concentration of a stress-regulating molecules.  Vice versa, the concentration evolves dynamically in response to surface flows and deformations. The numerical method is based on a coupling of surface and bulk equations in a finite element framework with deforming numerical grids \cite{Mokbel2020-uc, De_Kinkelder2021-vp}. 
Special care is taken to ensure that the surface mass and grid spacing are conserved under the occurring strong tangential flows. 
We validate the numerical results for spherical geometries with the linear stability analysis from \cite{Mietke2019-qf}. We  further explore the non-linear regime with a focus on ring-constrictions leading to cell division and geometrical effects for non-spherical geometries.

% Methods
\section{Model} \label{sec:model}
We consider a time-dependent surface $\Gamma\subset\mathbb{R}^3$ which encloses a fluid domain $\Omega_1$ and is surrounded by an external domain $\Omega_0$. 
The surface normal ${\bf n}$, pointing into $\Omega_0$, can be used  to define the surface projection operator $P_\Gamma = I-{\bf n}\otimes{\bf n}$, where $I$ is the identity matrix. 
The force balance on the surface reads 
\begin{align}
-\nabla_\Gamma\cdot S_\Gamma &= (S_1-S_0)\cdot {\bf n}  &\text{on}~\Gamma, 
\label{surface force balance}
\end{align}
where $\nabla_\Gamma\cdot$ is the surface divergence, $S_\Gamma$ is the surface stress tensor and $S_i$ for $i=0,1$ denotes the stress in the surrounding domains. % and $[\cdot]_0^1=\cdot_1-\cdot_0$ is the jump operator. 
The surface stress tensor contains contributions from surface viscosity and active and passive surface tensions \cite{Bothe2010-tz, De_Kinkelder2021-vp} %and bending stiffness
\begin{align}
S_\Gamma &= (\eta_b-\eta_s)\nabla_\Gamma\cdot {\bf v} P_\Gamma +2\eta_s D_\Gamma + \gamma P_\Gamma + S_{\Gamma, {\rm act}}&\text{on}~\Gamma.
\label{surface stress}
\end{align}
Here, $\eta_b$ and $\eta_s$ are the bulk and shear viscosity of the surface, respectively, $D_\Gamma = \frac{1}{2} P_\Gamma (\nabla_\Gamma \mathbf{v} + (\nabla_\Gamma \mathbf{v})^T)P_\Gamma$ is the surface strain rate tensor, $\gamma$ is a passive constant surface tension and $S_{\Gamma, {\rm act}}$ denotes the active tension to be specified later. 
In contrast to \cite{Mietke2019-ju,Mietke2019-qf}, the velocity field ${\bf v}$ contains both, tangential and normal parts such that Eqs.~\eqref{surface force balance}-\eqref{surface stress} include the force balance in  all directions. 
For simplicity, we neglect the bending stiffness of the cortical surface here, although included in our numerical framework. In \cite{Mietke2019-ju} it was found that the mechanochemical instability is independent of bending rigidity.

We describe the bulk domains on both sides of the active surface as passive incompressible Stokes fluids with a viscosity $\eta_i$ in $\Omega_i$ for $i=0,1$. Hence, the bulk stresses are
\begin{align}
S_i&=\eta_i(\nabla{\bf v}+\nabla{\bf v}^T)-p_i I &\text{in}~\Omega_i.
\label{bulk stress}
\end{align}
Here, $p_i$ are the hydrostatic pressures in the surrounding domains, distinguished here for $i=0,1$ because of the discontinuity of pressure at the surface. On the other hand, the velocity field ${\bf v}$ is assumed to be continuous because of the no-slip boundary condition at the cell surface. Balance of mass and forces lead to the Stokes equations 
\begin{align}
\nabla\cdot {\bf v} &= 0 &\text{in}~\Omega_i,\\
-\nabla\cdot S_i &= 0 &\text{in}~\Omega_i. 
\label{stokes}
\end{align}

The bulk and surface fluid equations are coupled to an advection-diffusion equation for the local concentration $c$ of stress generating surface molecules, e.g., myosin motor proteins:
\begin{align} \label{eq:concentration eq}
\partial_t^\bullet c + c\nabla_\Gamma\cdot{\bf v} &= D\Delta_\Gamma c - k_\text{off} c + k_\text{on} \bar{c} &\quad \text{on} \ \Gamma,
\end{align}
The first term represents the surface material time derivative of c, $\partial_t^\bullet c=\partial_t c+{\bf v}\cdot\nabla_\Gamma c$, which models the temporal change due to advection of $c$ along the surface. The second term accounts for changes in $c$ through local expansion or contraction of the surface. 
Area changes due to normal movement of curved surface regions are also included, since ${\bf v}$ contains the normal components of the velocity.
The first term on the right-hand side models the surface diffusion, with diffusion constant $D$ and Laplace-Betrami operator $\Delta_\Gamma$. The last two terms describe detachment with rate $k_{\rm off}$ from the surface into the bulk and attachment with rate $k_{\rm on}$ to the surface from the cytosolic concentration $\bar{c}$.

Finally, the system of equations is closed by the mechanochemical feedback of surface concentration to the fluid mechanics.
As in \cite{Mietke2019-ju, Mietke2019-qf} we consider an active surface tension $S_{\Gamma,{\rm act}} = \xi f(c) P_\Gamma$ which incorporates a contractility of the surface, scaled by $\xi$ and regulated by the concentration of motor proteins. Following previous work  \cite{Mietke2019-ju, Mietke2019-qf}, the Hill function $f(c) = \frac{2c^2}{c_0^2 + c^2}$ describes active tension as a monotonously increasing function of $c$. 
The force contribution of the active tension in Eq.~\eqref{surface force balance} can be rewritten into the form 
\begin{align*}
\nabla_\Gamma \cdot S_{\Gamma,{\rm act}} = \xi f(c) H{\bf n} + \xi f'(c) \nabla_\Gamma c,
\end{align*}
 where $H$ is the total curvature of the surface. The last term in this form, also called Marangoni term (e.g., \cite{Mokbel2017-zy}), represents a tangential force towards regions of high surface concentration and is the main driving force of the system. The induced flows transport proteins along the surface, thereby enriching spots of high concentration further. The resulting positive feedback loop gives rise to mechanochemical pattern formation. 
The process is accompanied by dynamic changes of surface shape. 
Assuming impermeability of the cell surface in the considered time scales, the movement of the surface and bulk domains is described by the normal part of the flow velocity, in particular $\partial_t \Gamma = {\bf n}~{\bf v}\cdot {\bf n}$. 

We nondimensionalize the equations using the diffusive timescale $\tau_D := \frac{R^2}{D}$, characteristic length scale $R$ (the initial cell radius) and the characteristic concentration of the equilibrium of surface/bulk protein exchange $c_0=\frac{k_\text{on}}{k_\text{off}}\bar{c}$.
The complete governing equations, now in non-dimensional variables, reduce to:
\begin{align}
-\nabla_\Gamma\cdot S_\Gamma &= (S_1-S_0)\cdot {\bf n} &\text{on}~\Gamma, \label{eq:nondim1}\\
S_\Gamma &= (1-\nu)\nabla_\Gamma\cdot {\bf v} P_\Gamma +2\nu D_\Gamma + Pe (\tilde{\gamma} + f(c))P_\Gamma &\text{on}~\Gamma, \label{eq:nondim2}\\
\nabla\cdot {\bf v} &= 0 &\text{in}~\Omega_i,\label{eq:nondim3}\\
-\nabla\cdot S_i &= 0 &\text{in}~\Omega_i, \label{eq:nondim4}\\
S_i&=\frac{\eta_i R}{\eta_b} (\nabla{\bf v}+\nabla{\bf v}^T)-p_i I &\text{in}~\Omega_i,\label{eq:nondim5}\\
\partial_{t}^\bullet c + c \nabla_\Gamma\cdot \mathbf{v}  &=  \Delta_\Gamma c - {\tau_D k_\text{off}} (c - 1) &\text{on}~\Gamma, \label{eq:nondim6}
\end{align}

with the  five non-dimensional model parameters $\tau_D k_\text{off}$, Péclet number $Pe=\frac{R^2 \xi}{D\eta_b}$, $\nu = \frac{\eta_s}{\eta_b}$, $\frac{\eta_i R}{\eta_b}$, and $\tilde{\gamma}=\frac{\gamma}{\xi}$.
Consistently with \cite{Mietke2019-qf}, we define the hydrodynamic length scale $L_h := \eta_b/\eta_1$ as a measure for surface viscosity. %, and additionally the viscosity ratio $L^\prime_h := \eta_b/\eta_1$.
%maybe invent a nondimensional number for $\eta R/\eta_b$

\section{Numerical Method}
We end up with a coupled system of surface and bulk hydrodynamics and surface concentration, on time-dependent domains. 
The stable discretization of such a system is a challenging task for which we develop a finite element scheme in the following. The scheme is based on connected numerical grids for the bulk domains, which share the same grid points with the surface grid.

\subsection{ALE approach for grid movement}
We use the Arbitrary Lagrangian-Eulerian (ALE) method to discretize the moving numerical grids. Thereby, we introduce the grid velocity ${\bf w}(x,t)\in\mathbb{R}^3$ which is used to update positions of grid points $x$ in the bulk domains and on the surface. 
While the surface $\Gamma$ needs to move with the flow in normal direction, tangential motion must be chosen differently such that the strong tangential flows do not deteriorate the grid quality. We follow the approach of \citep{De_Kinkelder2021-vp} and use 
\begin{align}
{\bf w}&={\bf n}~{\bf v}\cdot{\bf n} + P_\Gamma\cdot {\bf v}_{\rm avg} & {\rm on}~\Gamma, \label{eq:grid_velocity}
\end{align}
where ${\bf v}_{\rm avg}$ is some average velocity of the surface, e.g., the velocity of the cell barycenter. 
This form not only moves the surface in the correct way, ${\bf w}\cdot{\bf n} = {\bf v}\cdot{\bf n}$, but also ensures an even distribution of grid points along the surface. For example, for a purely tangential flow it guarantees ${\bf w}=0$, and for a purely translational flow (${\bf v}= {\bf v}_{\rm avg}$), it ensures ${\bf w}={\bf v}_{\rm avg}$.

%	\item ${\bf w}_\Gamma={\bf v}$
%	\item ${\bf w}_\Gamma={\bf n}~{\bf v}\cdot{\bf n}$

To keep a proper shape of the mesh, the grid velocity of the bulk domains is computed as a continuous harmonic extension of the surface grid velocity. In this work, the grid velocity $\textbf{w}$ is calculated in the fluid domain by solving the Laplace problem
\begin{align}
\begin{invisibleCases}
\Delta\textbf{w}&=0 &\text{in }\Omega_i, i=0,1, \\
%\textbf{w}&=\textbf{w}_\Gamma,~
\textbf{w}&={\bf n}~{\bf v}\cdot{\bf n} + P_\Gamma\cdot {\bf v}_{\rm avg}&\text{on }\Gamma, \\
\textbf{w}&=0&\text{on }\partial\Omega_i\setminus\Gamma, i=0,1.
\end{invisibleCases}
\label{eq:laplace}
\end{align}
Alternatively, other methods can also be used to extend the grid velocity into the fluid domain, for example by penalizing length changes of triangles/tetrahedra, see \cite{Mokbel2020-uc} for more details.

To obtain a consistent discretization, the calculated grid velocity $\textbf{w}$ is then subtracted in the convective term of the concentration equation. Therefore, the material time derivate $\partial_t^\bullet c$ on the surface is replaced by
\begin{align*}
\partial_{t|\hat{x}}  +({\bf v}-\textbf{w})\cdot\nabla_\Gamma ~,
\end{align*}
where $\partial_{t|\hat{x}}$ defines the time derivative along a quantity on a moving grid point.

\subsection{Space and time discretization}
\label{discretization}
\begin{figure}[ht]
    \centering
    \includegraphics[width=0.9\textwidth]{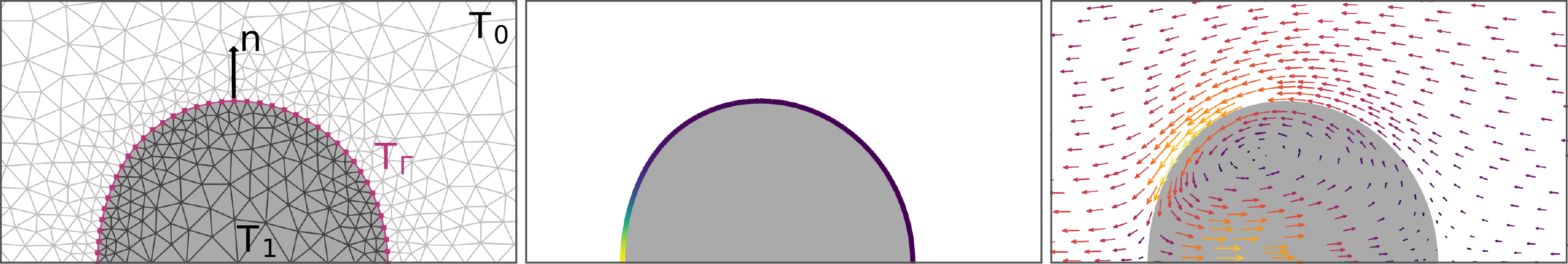}
    \caption{
    {\bf Computational Domain and Segregated Solver Steps:}
    {\bf left}
    Exemplary triangulation of the computational domain with $T_1$ (gray), $T_0$ (white) and $T_\Gamma$ (purple). 
    {\bf middle}
    Sub-step for the solution of the concentration on $T_\Gamma$. 
    {\bf right}
    Solution sub-step solving the hydrodynamics in $T_0$ and $T_1$. 
    }
    \label{fig:domain}
\end{figure}

In the finite element approach, we consider the separated triangulations $T_0$ of $\Omega_0$ and $T_1$ of $\Omega_1$, whereby the membrane $\Gamma$ that connects the two domains is triangulated by $T_{\Gamma}=T_0\cap T_1$. In particular, the connection is ensured by the fact that every grid point on the interface of $\Omega_0$ has a corresponding grid point on the interface of $\Omega_1$, sharing the same point coordinates. This allows us to enforce continuity of velocity across $\Gamma$, as well as to implement the jump conditions of the stress there. We present an example for the numerical mesh in \cref{fig:domain} on the left. 

The problem is discretized in time with equidistant steps of size $dt$. At each time step $n$, the grid is first updated, i.e., each grid point $x$ is  moved by $dt\cdot{\bf w}^{n-1}(x)$. Afterwards, the complete system is decoupled into the three subsystems, which are solved subsequently on the fixed new grid geometry. First, the equation for the molecule concentration is solved on the surface (cf \cref{fig:domain} in the middle panel). Afterwards the velocity field is computed from the coupled surface and bulk hydrodynamics (cf \cref{fig:domain} in the right panel), and in a last step, the grid velocity ${\bf w}$ is computed as explained in the previous section.

\subsubsection{Mass conserving discretization of molecule concentration}
\label{discretization concentration}
The active tension leads to tangential flows, which are the main driving force of the  surface hydrodynamics. 
Even in equilibrium situations with static surface shape, significant transport along the surface happens and is balanced by surface diffusion and exchange with the bulk concentration. 
Accordingly, it is crucial that the employed discretization ensures conservation of molecule mass even under strong tangential flows. 
The concentration equation guarantees conservation in the sense that surface molecule mass only changes by attachment and detachment of molecules:

\begin{align}
\frac{d}{dt}\int_\Gamma c 
&= - \int_\Gamma \tau_D k_\text{off} (c-1), \label{eq: mass conserv}
\end{align}
which follows from the fact that $\Gamma$ is a \textit{closed} surface and using the transport theorem for evolving hypersurfaces \cite[Thm.32]{Barrett2020-gy} with global surface parameterization induced by the velocity field ${\bf w}$. 
This conservation property can be carried over to the discrete level as follows. 
The ALE approach leads to the surface concentration system 
\begin{align}
\partial_{t|\hat{x}} c +({\bf v}-{\bf w})\cdot\nabla_\Gamma c + c\nabla_\Gamma \cdot {\bf v} &= \Delta_\Gamma c - \tau_D k_\text{off} (c-1) &\quad \text{on}~\Gamma.
\end{align}
Using that $({\bf v}-{\bf w})\cdot\nabla_\Gamma c = \nabla_\Gamma \cdot (({\bf v}-{\bf w}) c) -c\nabla_\Gamma \cdot {\bf v} + c\nabla_\Gamma \cdot {\bf w}$ we can rewrite the equation to 
\begin{align}
\partial_{t|\hat{x}} c + \nabla_\Gamma \cdot \left(({\bf v}-{\bf w})c\right) +c\nabla_\Gamma \cdot{\bf w} &= \Delta_\Gamma c - \tau_D k_\text{off} (c-1) &\quad \text{on}~\Gamma.
\end{align}
This formulation is amenable to a mass conserving discretization for tangential flows. An implicit Euler method leads to the time discrete scheme 
\begin{align}
\frac{c^n-c^{n-1}_{\hat{x}}}{dt} + \nabla_\Gamma \cdot \left(({\bf v}^{n-1}-{\bf w}^{n-1})c^n\right) +c^n\nabla_\Gamma \cdot{\bf w}^{n-1} &= \Delta_\Gamma c^n - \tau_D k_\text{off} (c^n-1) &\quad \text{on}~\Gamma,
\label{concentration time discrete}
\end{align}
where the subscript denotes the time step index and $c^{n-1}_{\hat{x}}$ is the concentration of the previous time step, $c^{n-1}$, but after the mesh update, i.e. the grid point coordinates have been
moved without changing the concentration values.

\noindent
The equation is discretized in space using piecewise linear surface functions,  
\begin{align}
C_{h}&:=\left\lbrace c\in C^0\left(\Gamma\right) \bigg \vert c_{\vert k}\in P_1\left(k\right),~k\in T_\Gamma \right\rbrace,
\end{align}
which leads to the weak problem: Find $c^n\in C_h$ such that for all $q\in C_h$,
\begin{align}
0&= \int_\Gamma \frac{c^n-c_{\hat{x}}^{n-1}}{dt} q - c^n({\bf v}^{n-1}-{\bf w}^{n-1})\cdot \nabla_\Gamma q + c^n q \nabla_\Gamma \cdot {\bf w}^{n-1} + \nabla_\Gamma c^n \cdot \nabla_\Gamma q + \tau_D k_{\rm off}(c^{n}-1)q ~d\textbf{x}.  \label{eq:conc_weakform}
\end{align}

\noindent
The mass conservation property of this formulation can be shown by testing this equation with $q=1$, which is clearly in $C_h$. We obtain
\begin{align}
0&= \int_\Gamma \frac{c^n-c_{\hat{x}}^{n-1}}{dt} + c^n \nabla_\Gamma \cdot {\bf w}^{n-1} + \tau_D k_{\rm off}(c^{n}-1) ~d\textbf{x}. 
\label{c equation tested with 1}
\end{align}
If the velocity along the surface is purely tangential, ${\bf v}\cdot {\bf n} = 0$ on $\Gamma$, it follows that ${\bf w}=0$ and the numerical grid remains stationary.
Hence, $c_{\hat{x}}^{n-1}=c^{n-1}$, and Eq.~\eqref{c equation tested with 1} reduces to
\begin{align*}
\int_\Gamma c^n ~d\textbf{x} -\int_\Gamma c^{n-1}  ~d\textbf{x} &= dt\int_\Gamma \tau_D k_{\rm off}(c^{n}-1) ~d\textbf{x},
\end{align*}
which is the discrete equivalent of the mass conservation property, Eq.~\eqref{eq: mass conserv}.

\subsubsection{Surface and bulk hydrodynamics}
\label{discretization hydrodynamics}
After one step of the surface concentration problem, the coupled equations of bulk and surface hydrodynamics are solved in a monolithic system. 
To this end, we define the viscous surface force 
\begin{align}
    F_{\Gamma,{\rm visc}}=\nabla_\Gamma\cdot \left[(1-\nu)\nabla_\Gamma\cdot {\bf v} P_\Gamma +2\nu D_\Gamma\right].
\end{align}
To stabilize strong contribution from surface viscosity, we discretize in time using a relaxation scheme with relaxation constant $\omega\in(0,1]$,
\begin{align}
F_{\Gamma,{\rm visc}}^{n} &= (1-\omega) F_{\Gamma,{\rm visc}}^{n-1} + \omega \nabla_\Gamma \cdot \left[(1-\nu)\nabla_\Gamma\cdot {\bf v}^{n-1} P_\Gamma +2\nu D_\Gamma^{n-1}\right] &\text{on}~\Gamma, \label{surface stress time discrete}
\end{align}
starting with $F_{\Gamma,{\rm visc}}^{0}=0$ in the first time step ($n=1$). 
%Time stepping errors which stem from the relaxation can be expected to appear at a timescale of $dt / \omega$, which we choose well below the typical timescale of the observed phenomena. 
Note, that all occurrences of $\Gamma$ here describe the current grid  $\Gamma^n$ and all involved quantities are assumed here to be moved onto this grid.
The coupled bulk surface system is discretized in time as,
\begin{align}
- F_{\Gamma,{\rm visc}}^{n} -Pe\nabla_\Gamma\cdot \left(\left(\tilde{\gamma} + f(c^n)\right)P_\Gamma\right) &= (S_1^n-S_0^n)\cdot {\bf n} &\text{on}~\Gamma, \label{bc stress discrete}\\
\nabla\cdot {\bf v}^n &= 0 &\text{in}~\Omega_i, \label{incompressibility time discrete}\\
-\nabla\cdot S_i^n &= 0 &\text{in}~\Omega_i, \label{bulk stress balance time discrete} 
\end{align} 
with
\begin{align}
S^n_i&=\frac{\eta_i R}{\eta_b} (\nabla{\bf v}^n+(\nabla{\bf v}^n)^T)-p_i^n I &\text{in}~\Omega_i.
\end{align}
{%\color{red}
Note, that an explicit coupling of surface evolution and  hydrodynamics typically leads surface tension related time step restrictions \cite{popinet2018numerical,denner2022breaching}. However, using the above time discretization we do not experience such issues. }This might be explained by either (i) the stabilizing effect of the additional surface viscosity, or (ii) the fact that surface viscosity itself induces time step restrictions which we mitigate (but not eliminate) by the above relaxation mechanism. 

For discretization in space we adopt the approach from \cite{Mokbel2020-uc} and introduce the following finite element spaces:
\begin{align}
\begin{invisibleCases}
V_h&:=\left\lbrace v\in C^0\left(\overline{\Omega}\right)\cap H^1\left(\Omega\right)\bigg \vert v_{\vert k}\in P_2\left(k\right),~k\in T_0\cup T_1\right\rbrace\\
P_{h,i}&:= \left\lbrace p\in C\left(\overline{\Omega}_{i}\right)\cap L^2\left(\Omega_{i}\right) \bigg \vert p_{\vert k}\in P_1\left(k\right),~k\in T_{i} \right\rbrace,~i=0,1.
\end{invisibleCases}
\label{eq:sobolev}
\end{align} 
Here, $V_h$ is the finite element space for the components of the velocity $\textbf{v}$. The usage of two separate spaces, $P_{h,i}, i=0,1$ ensures good approximation properties despite the pressure discontinuity across $T_{\Gamma}$.
The use of standard finite element spaces for the discretization of the discontinuous pressure leads to poor numerical properties, with an approximation order of only $\mathcal{O}(\sqrt{h})$ w.r.t. to the $L^2$ norm \cite{Gros2007-ht}.
Accordingly, the usage of two separate spaces, $P_{h,i}, i=0,1$, extends the standard Taylor-Hood finite element space by additional degrees of freedom of the pressure at the interface, such that the discontinuity can be exactly resolved. 

The weak form of the hydrodynamic equations then reads:
$ \text{Find }\left(\textbf{v}^n,p_0^n,p_1^n\right)\in V_h^d\times P_{h,0}\times P_{h,1}$ such that for all $\left(\textbf{z},q_0,q_1\right)\in V_h^d\times P_{h,0}\times P_{h,1}:$
\begin{align}
0&=\int_{\Omega_0}\frac{\eta_0 R}{\eta_b}\left(\nabla\textbf{v}^n+(\nabla\textbf{v}^n)^T\right):\nabla\textbf{z} - p_0^n\nabla\cdot\textbf{z} +q_0 \nabla\cdot{\bf v}^n ~d\textbf{x} \nonumber\\
 &+\int_{\Omega_1}\frac{\eta_1 R}{\eta_b}\left(\nabla\textbf{v}^n+(\nabla\textbf{v}^n)^T\right):\nabla\textbf{z} - p_1^n\nabla\cdot\textbf{z} +q_1 \nabla\cdot{\bf v}^n ~d\textbf{x} \nonumber\\
 &+\int_\Gamma F_{\Gamma,{\rm visc}}^{n}\cdot {\bf z} + Pe\nabla_\Gamma\cdot \left(\left(\tilde{\gamma} + f(c^n)\right)P_\Gamma\right) \cdot {\bf z}  ~d\textbf{x}, \label{eq:hydro_weakform}
\end{align} 
%vorzeichen stimmen
where Eq.~\ref{bc stress discrete} was directly incorporated as a natural boundary condition, after integration by parts has been used on the bulk stress terms.
Details on the implementation in the finite-element framework AMDiS \cite{Vey2007-af,Witkowski2015-yz} and {\color{black} the numerical convergence study can be found in the appendix.
The presence of spurious currents is found to be negligible (magnitudes $\sim 10^{-8}$).
}

% Results
\section{Results}
\label{sec:results}

In this section, we start by 
{\color{black}
presenting full three-dimensional simulations of the coupled system. We show that these lead to formation of stationary concentration patterns, namely a single spot or a ring of high concentration.  
Since those two patterns are the most biological relevant and both are axisymmetric, we exploit the rotational symmetry and use an axisymmetric model in the subsequent sections. 
Details on the implementation of the axisymmetric numerical method can be found in the Appendix.
With the axisymmetric model, we compare
}
our new numerical model to the linearized analytical results of \cite{Mietke2019-qf} and further investigate the non-linear behavior of the system.
%{\color{black} \st{Details on the implementation of the axisymmetric numerical method can be found in the Appendix.}}
If not stated otherwise, each simulation is started on a spherical surface geometry covered with an initial concentration $c(x)=c_0(1+\epsilon(x))$, where $\epsilon(x)\in [- 10^{-5}, 10^{-5}]$ denotes  uniformly disturbed random noise on each grid point $x$, with $\int_\Gamma \epsilon(x) \, dx = 0$.
The default parameters for the following simulations are given in \cref{tab:params}.
The ratio between the inner and the outer fluid viscosity is set such that the outer fluid has negligible influence, and we consider only the active surface tension contribution by setting  $\tilde{\gamma} = 0$. 

\begin{table}[!th]
    \centering
    \begin{tabular}{ccl}
        Symbol & Value & Meaning\\
        \hline
        $\pe$ & 150 & Péclet number \\
        $L_h / R$ & 1 & Hydrodynamic length scale \\
        $\tau_d k_\text{off}$ & 10 & Non-dimensional attachment / detachment rate  \\
        $\nu$ & 1 &  Ratio of surface shear to surface bulk viscosity\\
        $\tilde{\gamma}$ & 0 & Ratio of passive to active surface tension \\
        $\eta_0 / \eta_1$ & $\num{1e-2}$ & Ratio of outer to inner bulk viscosity\\
        \hline
        $dt / \tau_D$  &  $\num{1e-4}$ & time step size\\
        $h_\Gamma /  R$ &  $\num{8e-2}$ & max mesh size on $\Gamma$ \\
        $\omega$ & $\num{1e-1}$ & Relaxation parameter for the surface viscosity stress
    \end{tabular}
    \caption{Default model and simulation parameters}
    \label{tab:params}
\end{table}

{\color{black}
\subsection{Pattern Formation in 3D}
\label{sec:3d}
}
Consistent with \cite{Mietke2019-qf}, we find that the system gives rise to pattern formation if the activity (measured by the Péclet number $Pe$) is sufficiently high. In this case, a positive feedback loop sets in: regions of higher surface concentration are contracted, thereby inducing a flow which transports more contractile molecule towards this region. 
This feedback loop is finally balanced when advected and diffusive transport cancel each other out, leading to a stationary concentration pattern along the surface and a persistent flow towards  the concentration peaks.
Linear stability analysis in \cite{Mietke2019-qf} suggests parameter ranges where a ring pattern and single spot pattern are possible. 

{\color{black}
\Cref{fig:3D} shows full three-dimensional simulation results for two parameter sets where a ring pattern and a single spot are predicted.  
In both cases, the initial random concentration field is the same. 
Due to the different values for the Péclet number~$\pe$ and hydrodynamic length scale~$L_h / R$, the time evolution, shape, and final stationary configuration differ.
For $\pe = 91.7$ and $L_h / R = 0.3$, a ring pattern emerges (\cref{fig:3D_L2} and supplemental movie 1), while for $\pe = 50.8$ and $L_h / R = 0.63$ a single high concentration spot emerges  (\cref{fig:3D_L1} and supplemental movie 2). For the choice of parameters, see Sec.~\ref{sec:validation}.
Both stationary configuration are rotationally symmetric. The axis of rotation varies in each simulation depending on the given random initial data. 

Interestingly, before the final ring pattern emerges, two concentration spots opposite each other emerge (see \cref{fig:3D_L2} at $t / \tau_D = 1$). 
These two spots dissolve into the final ring pattern as seen at $t / \tau_D = 2$ and $t / \tau_D = 4$.
The cell becomes slightly constricted, and after $t / \tau_D = 4$, the ring and the cell shape stay stationary. 

We note that we find similar patterns (rings and single spots) for other random initial conditions. In particular, all observed patterns converge into an axisymmetric stationary state. This motivates to exploit the rotational symmetry and use an axisymmetric implementation (see Appendix) in the subsequent sections. 

%The time point where the instability occurs, and the pattern is clearly distinct, depends on the parameter sets.  For the case of the ring pattern in \cref{fig:3D_L2}, the instability occurs around $t / \tau_D \sim 1.25$. For the single spot in \cref{fig:3D_L1} around $t/\tau_D \sim 2.5$.

\begin{figure}
    \centering
    {\phantomsubcaption\label{fig:3D_L2}}%
  	{\phantomsubcaption\label{fig:3D_L1}}%
    \includegraphics[width=0.8\textwidth]{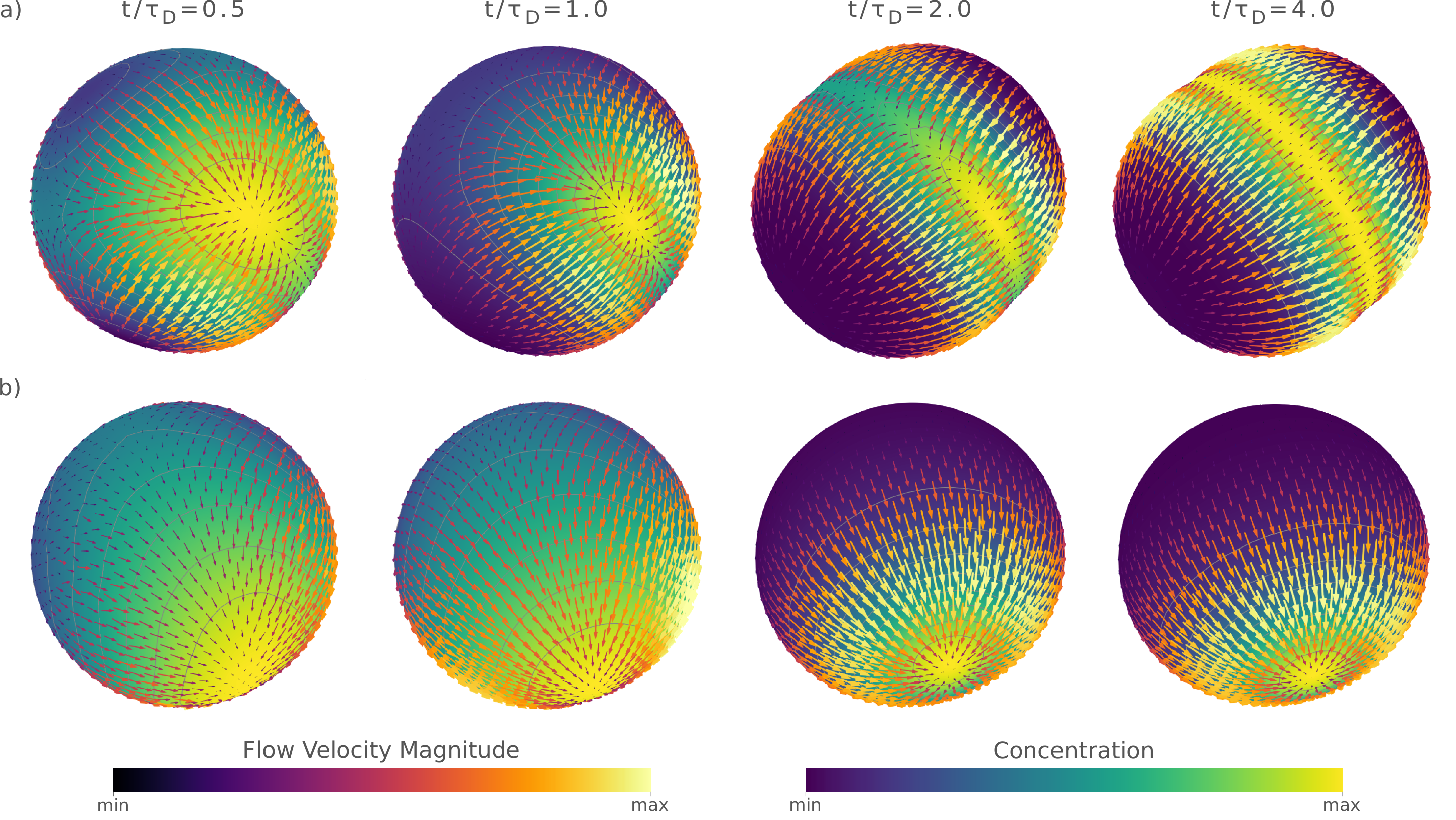}         
    \caption{
        {\color{black}
        {\bf Full Three-Dimensional Simulations of an Active Surface:}
        {\bf (a)} 
       Evolution of the concentration field and the tangential surface velocity for five different time points. 
       The parameters (see below) are set such that a concentration ring emerges, which becomes stationary after $t / \tau_D > 4$.
        {\bf (b)}
        Four different time points  for a parameter set where a single concentration peak emerges.
        The peak is stationary after $t / \tau_D > 3$.
        The concentration field and the tangential surface velocity indicated by the arrows are scaled for each time point, and parameter sets separately. 
        The grey lines are contours of the concentration field.
        Parameters:
        (a)
        $\pe = 91.7$,
        $L_h / R = 0.3$,
        (b):
        $\pe = 50.8$,
        $L_h / R = 0.63$,
        In both cases:
        %$\tau_D k_\text{off} = 10$,
        %$\nu = 1$,
        %$\tilde{\gamma} = 0$,
        $dt / \tau_D = \SI{5E-4}{}$,
        $h_\Gamma / R = \SI{2e-1}{}$,
        $\epsilon(x)\in [- 10^{-2}, 10^{-2}]$.
        %The outer bulk viscosity is $100\times$ smaller than the inner bulk viscosity.
       }
    }
    \label{fig:3D}
\end{figure}

} % end red

\subsection{Validation}
\label{sec:validation}
%{\color{black}\st{Consistent[$\hdots$] concentration peaks.}}
To validate our new model, we compare the stationary patterns {\color{black} of axisymmetric simulations} to the analytical results based on linear stability analysis of \cite{Mietke2019-qf}. 
There, the concentration distribution was quantified in terms of axisymmetric spherical harmonics modes 
$Y^0_\ell(\theta)$, where $\ell$ is the mode number and $\theta$ is the polar angle of a point on the surface. 
The critical Péclet number $Pe^\ast_\ell$ (the Péclet number above which patterns of mode $\ell$ emerge) was derived from linear stability analysis to be \cite{Mietke2019-qf}
\begin{align}
    \pecrit = \frac{1}{\cpfc} \left(1 + \frac{\taudk}{\ell(\ell+1)}\right) \times \left[ \ell(\ell+1) + \nu \left((\ell-1)(\ell+2) +(1+2\ell) \frac{R}{L_h} \right) \right]. \label{eq: critical PE}
\end{align}
We are particularly interested in the cases $\ell = 1$ (one single spot of high concentration at one of the poles) and the negative $\bar \ell = 2$ mode (ring around the cell leading to a cell constriction). 
%In \cite{Mietke2019-qf}\todo{check if correct publication}, the growth rate $\tau_D \lambda_\ell$ for %the different modes $\ell$ is given by 
%\begin{align}
%    \tau_D \lambda_\ell &= -l(l+1) \left(1 + \frac{\taudk}{l(l+1)} - \frac{\pe \cpfc}{(1+2l)\nu R/L_h + l(l+1)(1+\nu) - 2\nu}   \right)
%\end{align}

\begin{figure}
    \centering
    {\phantomsubcaption\label{fig:overview_phasediagram}}%
  	{\phantomsubcaption\label{fig:overview_maxconc}}%
  	{\phantomsubcaption\label{fig:overview_patterns}}
    \includegraphics[width=0.8\textwidth]{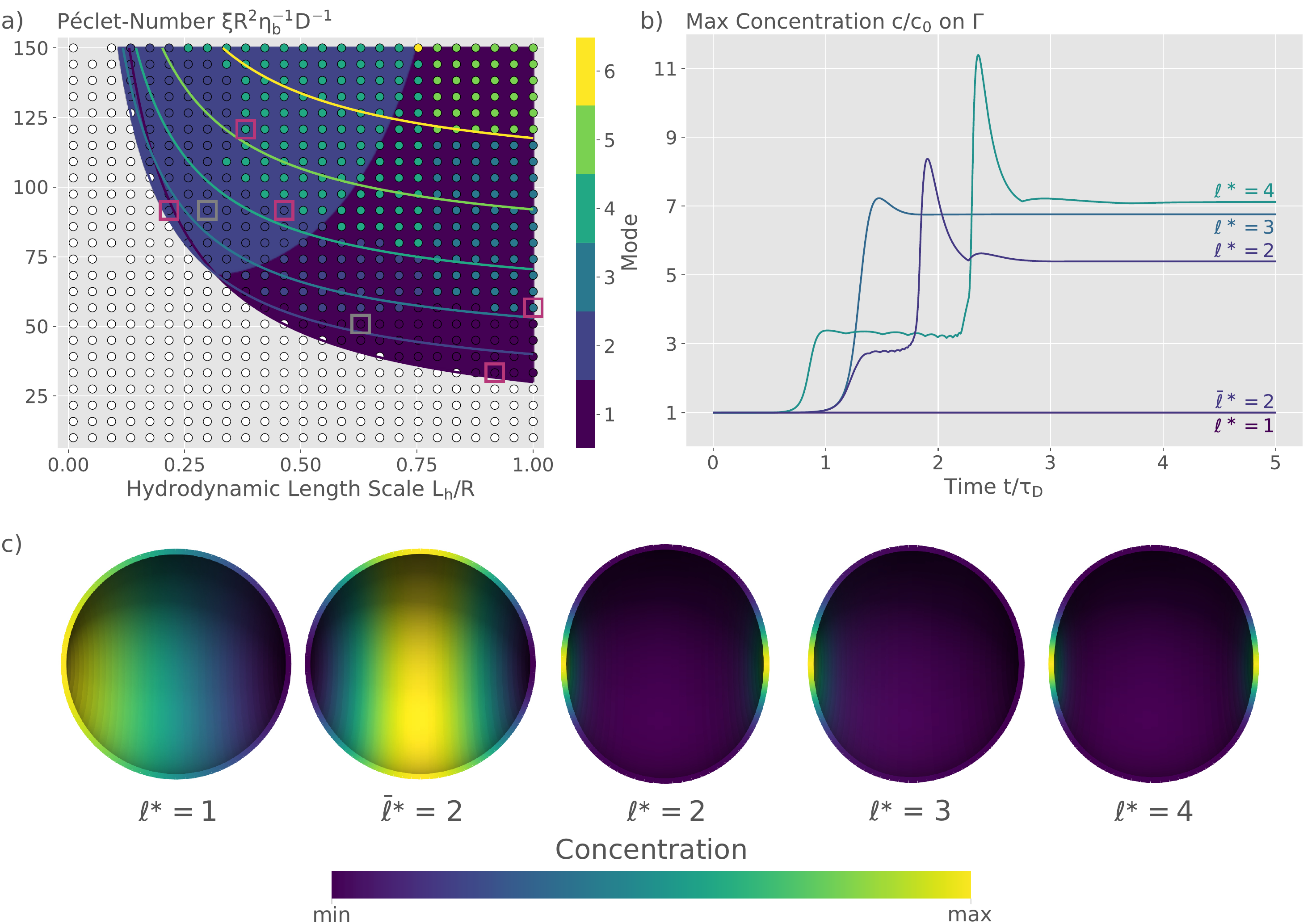}         
    \caption{
        {\bf Numerical Validation of obtained Stationary Patterns:}
        {\bf (a)} 
        Comparison of numerical results with linear stability analysis \cite{Mietke2019-qf} by classification of patterns in spherical harmonic modes. Numerically obtained stationary patterns (circles colored by $\ell^\ast$) emerge in the same parameter region as linear stability analysis predicts (background color). 
        White circles correspond to homogeneous concentration $c\approx c_0$, gray background color indicates the region where no pattern formation is expected \cite{Mietke2019-qf}. %The colour in the background corresponds to the mode with the maximal growth rate.
        The contour lines indicate the critical Péclet number $Pe_\ell^\ast$ for different modes. Numerical patterns with mode $\ell$ emerge only above the corresponding $Pe_\ell^\ast$ line. 
        {\bf (b)}
        The maximum concentration peak on the surface for the five simulations highlighted in (a) indicates that all configurations reach a steady state.
        {\bf (c)}
        Concentration distribution for the simulations highlighted by {\color{black}purple} squares in (a). The inverted $\bar\ell^\ast=2$ is located at the boundary of the transition region.
        {\color{black}The parameter sets used in \cref{fig:3D} are highlighted by grey squares.}
    }
    \label{fig:overview}
\end{figure}

In \cref{fig:overview_phasediagram} we conduct a comparison for the exemplary parameters presented in \cite{Mietke2019-qf}: $\taudk = 10$, $\cpfc = 1$ and $\nu = 1$, {\color{black}$L_h/R \in [0.01, 1]$, and $\pe \in [0, 150 ]$}. In this parameter region, the modes with a maximal growth rate are $\ell = 1$ and $\ell = 2$ (see filled areas in \cref{fig:overview_phasediagram}). 
%Higher order modes do start to grow too but not as fast as the first two. 
The lines indicate the critical Péclet number (Eq.~\eqref{eq: critical PE}) for each mode (modes only grow above this line).
The observed numerical patterns are categorized by the discrete Pearson correlation coefficient with the spherical harmonics functional $Y^0_\ell(\theta)$, defined by
\begin{align}
    r_\ell = \frac{\sum^N_{i = 0} (c_i - \overline{c}) Y^0_\ell(\theta_i)}{\sqrt{\sum^N_{i = 0} (c_i - \overline{c})^2 \sum^N_{i = 0} \left(Y^0_\ell(\theta_i) \right)^2}} \quad \text{for} \ \ell = 0, ..., 6,
\end{align}
where $c_i$ is the concentration on vertex $i$, $\overline{c} = \frac{1}{N} \sum^N_{i = 0} c_i$ the average concentration and $\theta_i$ the polar angle of $i$-th vertex. 
Concentration distributions with $\max_i{(c_i - \bar c)} \leq 10^{-5}$ are classified as stable (no pattern).  

Each circular marker in \cref{fig:overview_phasediagram} corresponds to one simulation. 
The dominant mode of each simulation is determined by $\ell^\ast := \argmax_\ell |r_\ell|$.
The transition from unstable (white markers) to emerging patterns (markers filled by the respective colors) coincides with the analytical prediction. 
%Exactly on the interface it would take an infinity amount of time for the instability to occur, thus for the few misclassified points the simulation end time $\hat{t} = 10$ is not enough. 
On the interface between the unstable (pattern) and stable (no pattern) regions, we obtain an almost perfect match regarding the predicted dominant modes. 
In this region, neither the geometry differs heavily from the initial sphere nor is the concentration distribution far away from $c_0$, such that linear stability prediction agrees with the fully nonlinear simulation. 
In the interior of the unstable phase, we do 
%not only get a more complex, initial concentration distribution dependent arrangements of $\ell = 1$ and $\ell = 2$ modes but also 
find higher order dominant modes. 
This results from the mixture of being away from the linear regime as well as geometrical effects due to shape changes.

The different times for the instability to occur is shown in \cref{fig:overview_maxconc} for the five simulations marked with purple squares in \cref{fig:overview_phasediagram}.
The simulations are chosen by the highest correlation coefficient $r_\ell$ for the first four modes plus the $\ell^\ast = 2$ mode at $L_h = 0.46$ and $\pe = 91.7$. 
For the $\ell = 1$ and $\ell = 2$ mode, the growth of the concentration peak is very small compared to the higher order modes ($c_{\rm max}=1.000054$ and $c_{\rm max}=1.000047$, respectively), which is due to the proximity to the phase transition (between stable and unstable) of the chosen parameters. 
The higher the Péclet number and / or the hydrodynamic length scale, the faster the instability occurs. 
All simulations reach an almost steady state where the surface concentration does not significantly change over time anymore.

In \cref{fig:overview_patterns} we show the resulting patterns, again for the five marked simulations. 
The asymmetric $\ell^\ast = 1$ (single spot) and the negative $\bar \ell^\ast = 2$ (concentration ring) are presumably the biologically most relevant modes. These are found near the transition region between mechanochemically stable and unstable phases.
The symmetric $\ell = 2$ mode is recovered too. 
Depending on the random initial contribution, the positive and negative $\ell = 2$ mode may occur for the same set of parameters. Similarly, for $\ell^\ast=1$, the concentration peak may appear in the left or the right pole. 
For higher order modes, the classification is non-trivial as the shape changes. Nevertheless, the Pearson correlation-based measure does reliably distinguish between symmetric and asymmetric concentration distribution. 

\subsection{Results in the non-linear regime} \label{sec:properties}

\begin{figure}
    \centering
    {\phantomsubcaption\label{fig:maxvelconc_conc}}%
  	{\phantomsubcaption\label{fig:maxvelconc_vel}}
    \includegraphics[width=0.8\textwidth]{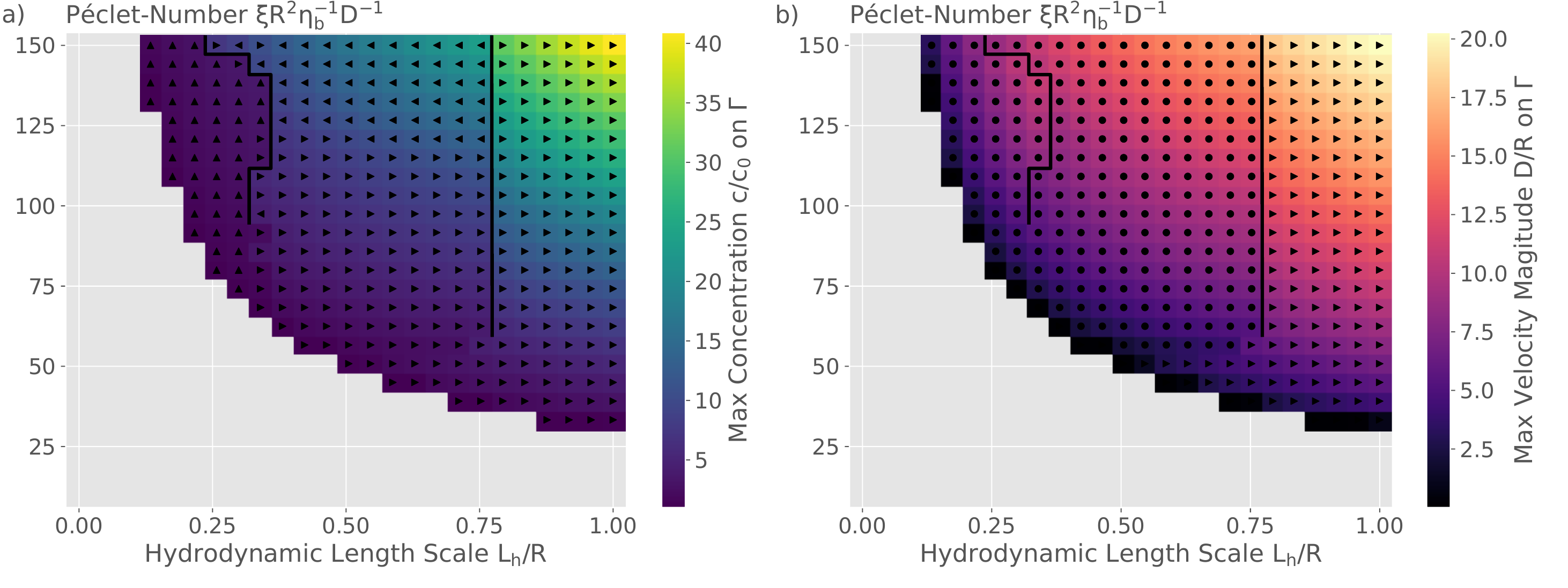} 
    \caption{
        {\bf Maximum Surface Concentration and Surface Velocity}
        {\bf (a)}
        Resulting maximum concentration peak on the surface of the phase diagram shown in Figure \ref{fig:overview}.         
        {\bf (b)}
        Maximum resulting flow magnitude on the surface. The parity of the resulting dominant mode is indicated by circles (even $\ell^\ast$) and triangles (odd $\ell^\ast$).
        Apparent discontinuities (indicated by curves) occur at the transition from $\ell^\ast=2$ to $\ell^\ast=4$ and from even to odd modes. 
        }
    \label{fig:maxvelconc}
\end{figure}

With the new numerical model at hand, we can investigate the non-linear dynamics away from the linear regime. 
In \cref{fig:maxvelconc}, the maximum surface concentration and surface velocity magnitude are shown. 
Both quantities increase  for higher Péclet number (which scales the activity $\xi$ of the system) and for more dominant surface viscosity (measured by $L_h$).
While the peak concentration increases superlinear in the Péclet number, the maximum velocity increases only sublinear. This is reasonable, as higher concentration peaks go along with a more dense localization of the concentration around its peak (data not shown). 
%Increasing the Péclet number is equivalent to increase the scaling factor $\xi$ of the active surface tension as $R_0$, $D$ and $\eta_s$ are kept constant \todo{use eta b everywhere instead of eta s}.
%On the other hand, increasing the hydrodynamic length scale $L_h=\eta_s/\eta_1$ has a similar effect. 
%The magnitude of the concentration peak directly influences the maximum velocity magnitude on the surface, as shown in \cref{fig:maxvelconc_vel}. 
The markers in \cref{fig:maxvelconc_conc} and \cref{fig:maxvelconc_vel} indicate whether the maximum Pearson correlation is found at an even or odd mode $\ell^\ast$. Even modes (circular markers) are more symmetric than odd modes, resulting in lower concentration peaks and thus lower velocity magnitudes on the surface.

Interestingly, there appears to be a clear discontinuity in the maximum concentration and maximum velocity around $L_h = 0.79$ (indicated by a black line). This transition goes along with the switch of $\ell^\ast$ from even (left) to odd (right), with lower velocities and lower concentrations for even $\ell^\ast$. 
A similar discontinuity is found around $L_h\approx 0.3$ in \cref{fig:maxvelconc_conc}, where all data points left of the black curve  show significantly lower concentration peaks. These data points correspond to $\ell^\ast=2$ (cf. Figure \ref{fig:overview}a).
We conclude that the mechanochemical instability seems, in general, much weaker for symmetric patterns (even modes) and in particular for a ring constriction $\ell^\ast=2$. 

\subsubsection{Ring Constrictions and Cell Division}
\label{sec:evenmodes}
\begin{figure}
    \centering
    {\phantomsubcaption\label{fig:evenmodes_radii}}%
  	{\phantomsubcaption\label{fig:evenmodes_slipping}}%
  	{\phantomsubcaption\label{fig:evenmodes_forced}}
    \includegraphics[width=0.8\textwidth]{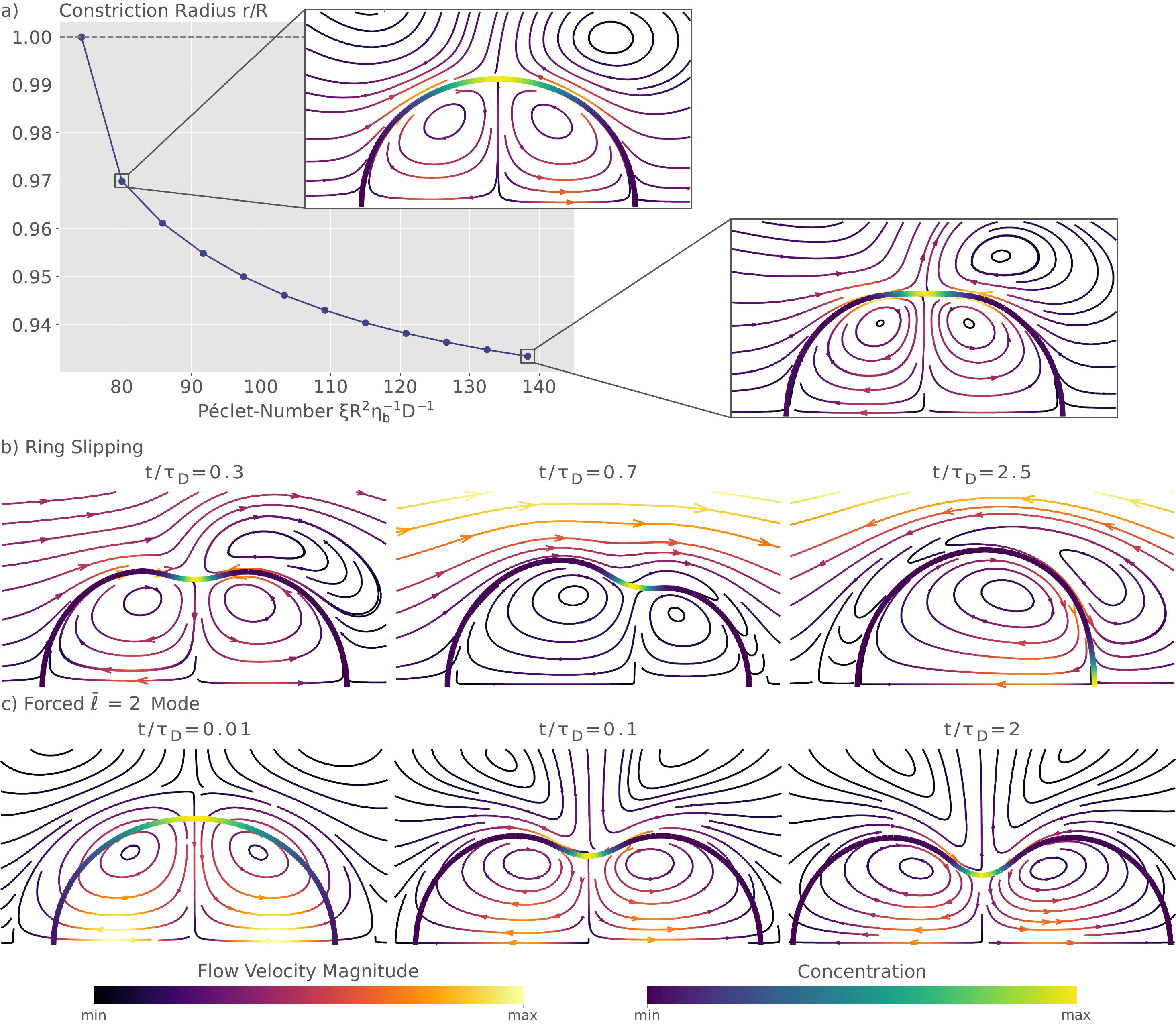}
    \caption{
    {\bf Cell Division by Contractile Ring Formation}
    {\bf(a)} 
    Size of the stationary constriction for the negative $\bar \ell^\ast = 2$ modes for $Pe \in [74.2, 138.3]$ and $L_h = 0.3$. 
    The extracellular flow is not fully symmetric due to the randomized initial concentration and the low outer viscosity. {\bf(b)} 
    Stronger ring constrictions emerge for high Péclet numbers and hydrodynamic length scales (here $Pe = 150, L_h = 1$) with minimal constriction radius of 0.8. 
    Such patterns are only transient, and the ring slips to either side (here around {$t / \tau_D= 0.7$}) leading to a stationary state without ring constriction. %$\ell^\ast = 5$ mode is reached.
    {\bf(c)} 
    Strong ring constrictions can be triggered by initializing the concentration field with a higher concentration in the center. Here for $Pe = 150$ and $L_h = 1.0$ the cell constricts significantly and would probably lead to cell division after {$t / \tau_D= 0.2$}. As such strong deformations deteriorate mesh quality, the simulation {\color{black} becomes} unstable at later times.
    Deviating parameters:
    (a)
    $L_h / R = 0.3$.
    (b):
    $\omega = \num{1e-2}$,
    $dt / \tau_D = \num{5e-5}$,
    $h_\Gamma / R = \num{4e-2}$,
    different random seed.
    (c), difference to (b):
    $\tau_d k_\text{off} = 0$, 
    $\eta_0 / \eta_1 = 1$,
    $dt / \tau_D = \num{2.5e-5}$,
    $\bar c \approx Y^2_0$, $\epsilon=0$.
    }
    \label{fig:evenmodes}
\end{figure}

The division of a mother cell into two daughter cells is the most radical shape change which a cell undergoes. 
This process is preceded by the symmetry-breaking formation of a ring of contractile molecules \cite{Mayer2010-ex,Pollard2010-ft}. 
The theoretical model of Mietke et al.\ 2019 \cite{Mietke2019-qf} predicts parameter regions where such a ring may emerge, indicated by a dominant $\ell = 2$ mode.
%The corresponding spherical harmonics is characterized by two symmetric peaks at the two poles of the sphere. 
%Inverted, called negative $\bar \ell = 2$ mode here, is a stable ring at the equator, the axis of which being defined by a spontaneous symmetry-breaking event. 
%The high concentration ring leads to a constriction of the sphere due to the increased surface tension, resembling the start of cell division.
Using the full numerical model, we observe such modes %the $\ell = 2$ and $\bar \ell = 2$ mode in our simulations,
even in regions where linear stability analysis predicts $\ell = 1$ to be the fastest growing mode (see \cref{fig:overview_phasediagram}). 
In these cases, the high concentration ring leads to a constriction of the cell due to the increased surface tension around its equator. The constricted radius is shown in \cref{fig:evenmodes_radii} for $L_h = 0.299$ and various Péclet numbers ($Pe \in [74.167, 138.333]$) which yield a ring pattern ($\bar \ell^\ast = 2$). 
The maximum attainable constriction is about 92\% (of $R$). This general behavior was found for various other parameter regimes and is consistent with the observation in Sec.~\ref{sec:properties}: The instability of ring formation and the associated shape changes seems weak compared to non-symmetric modes. 

In more unstable parameter regimes (higher $Pe$ and $L_h$), contractile ring formations were only transiently observed. An example is shown in \cref{fig:evenmodes_slipping}. There a contractile ring forms at early time ({$t / \tau_D= 0.3$}) and leads to a constriction of the cell to $0.8R$. But as non-symmetric modes seem to be more dominant in the non-linear regime, the contractile ring slips to one of the poles at later times ({$t / \tau_D \approx 0.7$}).
This exemplary behavior was found in all of the various cases which we tested in regimes of higher activity. 
Note that, such a ring slipping is also reported in vitro, e.g., in fission yeast cells \cite{Mishra2012-ej}.

Thus, we did not observe higher constriction leading to potential cell division, suggesting that there might be another biological mechanism involved. 
In \cite{Mietke2019-qf} the biasing signaling cues due to the mitotic spindle were included by imposing a higher local recruitment rate of proteins along the contractile ring. 
This goes along with a redefinition of the initial equilibrium concentration $c_0$. 
Similarly, we test the latter hypothesis by using an initial concentration which is localized in the cell center:  $c=1-\left|2\theta/\pi-1\right|$, which gives $c=1$ along the equator and c=0 in the poles. 
We conduct a simulation with higher hydrodynamic length scale, $L_h = 1$, and lower molecule exchange with the cytoplasm, $\tau_D k_{\rm off}=0$, such that the mechanochemical instability is more pronounced. 
The result, shown in \cref{fig:evenmodes}, indicates that a strongly localized concentration peak emerges in the center. The associated contractile flows lead to a significant constriction of the cell around {$t / \tau_D= 0.2$}. Over time, the neck formation accelerates and would probably lead to a break-up (cell division). However, shortly after  {$t / \tau_D= 0.2$}, the numerical grid becomes too distorted to accurately represent the cell shape and interior fluid. 
Thus, even if the dominant mode would favor an asymmetric pattern, the initial distribution leads to a symmetric ring constriction. It appears that the strongly evolved cell shape locks the high concentration peak in the center of the geometry.
Therefore, geometrical effects seem to play an important role in the dynamics of mechanochemical pattern formation. These effects are investigated further in Sec.~\ref{sec:geometric effects}.
%With our explicit representation of the cell surface it is not possible to simulate a full cell division and numerical instabilities occur shortly after the last time step in \cref{fig:evenmodes_forced}).

\subsubsection{Cell Polarization and Swimming}
\begin{figure}
    \centering
    \includegraphics[width=0.8\textwidth]{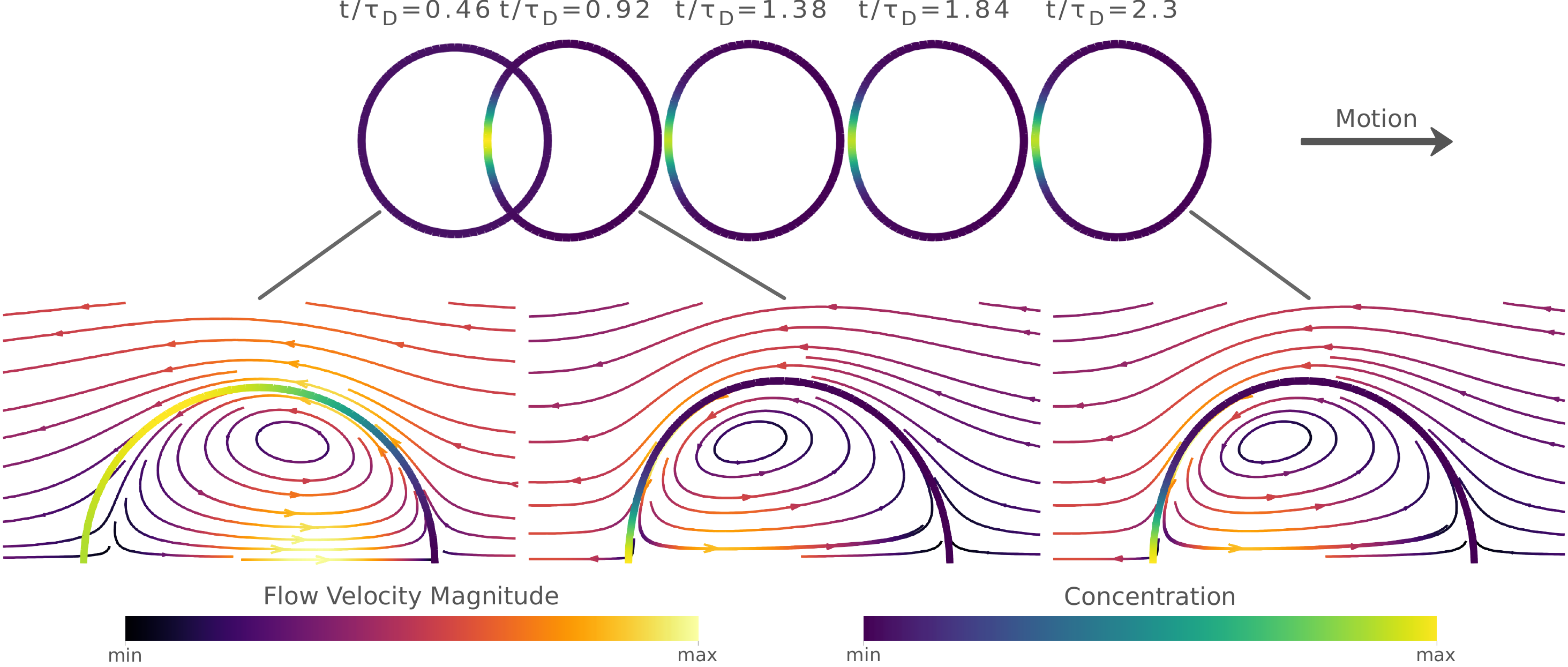}
    \caption{
    {\bf Swimming Cell:}
    {\bf (top)} Cell position and concentration distribution at 5 equidistant time points.
    %{\color{black}\st{indicated by $\hat t := t / \tau_D$}}. 
    Starting with a random initial distribution, the dynamic starts at around {$t / \tau_D= 0.46$} leading to a concentration peak at the left pole which induces motion to the right. Maximum concentration occurs around {$t / \tau_D= 0.92$}. A stationary configuration is reached, and the cell propels itself with constant velocity to the right.
    {\bf (bottom)}
    Concentration distribution and flow profiles (relative to cell motion) at the indicated time points. 
    Deviating parameters: 
    $\pe = 100$,
    $\eta_1 / \eta_0 = 1$,
    $\omega = \num{1e-2}$.
    }
    \label{fig:oddmodes}
\end{figure}

All left/right asymmetric patterns (indicated by odd $\ell^\ast$) induce asymmetric flow fields, which can set the cell into motion.
%The biologically most relevant modes are the $\ell = 1$ and $\overbline{\ell} = 2$ modes.
%The $\ell = 1$ mode (or more generally the asymmetric modes) leads to a translation of the cell through the surrounding fluid. 
In \cref{fig:oddmodes} we show this behavior for a parameter set which yields $\ell^\ast=1$. Starting from a random initial concentration,  the instability sets in around $t / \tau_D \sim 0.46$, reaches the highest concentration peak around {$t / \tau_D \sim 0.92$} and evolves into a stationary concentration distribution after {$t / \tau_D > 1.2$} (see \cref{fig:oddmodes} top). 
This goes along with a continuous flow towards the concentration peak (\cref{fig:oddmodes} bottom). 
The corresponding tangential acceleration along the cell periphery pushes the cell forward (mind the no-slip condition at the cell surface). Eventually, a stationary velocity is reached as the acceleration is balanced by the drag force in the surrounding medium. 
Note, that depending on the random initial distribution, the concentration peak may develop on the left or right side of the cell, leading to motion to the opposite side.

The observed continuous motion mediated by persistent myosin contraction at the rear is consistent to typical modes of cell crawling \cite{Lammermann2009-cd}. A similar swimming mechanism driven by Marangoni forces is also observed in the spontaneous self-propulsion of active colloids and droplets \cite{Zottl2016-hm, Whitfield2016-qu}

\subsubsection{Different Stationary States}
\label{sec:differentstates}
\begin{figure}
  	{\phantomsubcaption\label{fig:diffstates_cmax}}%
  	{\phantomsubcaption\label{fig:diffstates_kymograph}}
    \centering
    \includegraphics[width=0.8\textwidth]{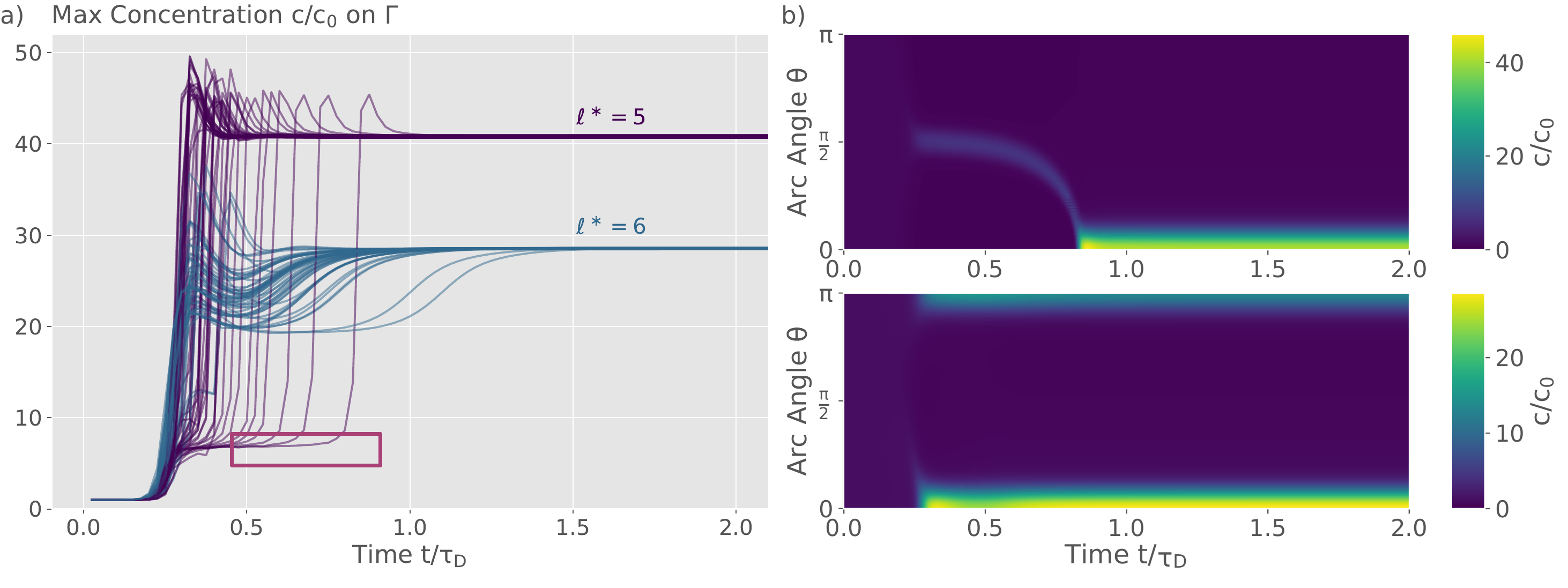}
    \caption{
    {\bf Several Stationary States}
    {\bf (a)}
    Maximum concentration on the surface $\Gamma$ for 100 different random seeds.
    Two different stationary deformations are reached, correlating with an $\ell = 5$ and $\ell = 6$ mode, some simulations assume a transient $\bar \ell^\ast = 2$ mode at the beginning (see purple marked region) similar to the slipping ring in Fig.\ref{fig:evenmodes}b. 
    {\bf (b)}
    Representative kymographs (concentration field over arc angle $\theta$ and time {$t / \tau_D$}) for the two stationary cases from (a) % random seed = 0 (bottom) and seed = 54 (top).
    Deviating parameters: 
    $\omega = \num{1e-2}$,
    $dt / \tau_D = \num{5e-5}$,
    $h_\Gamma / R = \num{4e-2}$.
    }
    \label{fig:diffstates}
\end{figure}

Close to the phase transition, linear stability analysis is able to compute growth rates of different surface modes, thus predicting the dominant mode which will outgrow other modes to generate a well-defined surface pattern \cite{Mietke2019-qf}. 
Interestingly, we observed that further away from the phase transition, stationary configurations are not unique anymore. %in regions of high Péclet numbers or large hydrodynamic length scale.
Instead, we did find several stationary configurations for the same parameter set.
In \cref{fig:diffstates} we show an example for 100 simulations with equal parameters, but each with different random seed $\epsilon(x)$. 
From those simulations, 46 reached an $\ell^\ast = 5$ mode and the remaining 54 an $\ell^\ast = 6$ mode. 
Several simulations reach a quasi-stationary state (see the purple square in \cref{fig:diffstates_cmax}). 
All of those simulations are ring patterns with $\bar \ell^\ast = 2$ but the concentration peaks slip away to one of the poles reaching a stationary $\ell^\ast = 5$ mode (see \cref{fig:diffstates_kymograph} top for the kymograph and \cref{fig:evenmodes_slipping} for the flow profiles). 
The kymograph of the configuration for the $\ell^\ast = 6$ is shown in \cref{fig:diffstates_kymograph} on the bottom.

To rule out that the occurrence of multiple stationary states is solely due to shape changes, we ran the same simulation on a stationary sphere, where we constrain the velocity field to 
\begin{align}\label{eqn:vn0}
    \mathbf{v} \cdot \mathbf{n} &= 0 & \text{on} \ \Gamma.
\end{align}
Even with these eliminated shape changes, we obtained the two different stationary configurations (data not shown). 
We conclude that several stationary configurations are possible for the fully non-linear system, at least for higher $Pe$ and $L_h$, i.e., far from the linear stability regime. 

\subsection{Geometrical Effects}
\label{sec:geometric effects}

Theoretical and experimental studies have suggested a strong coupling between membrane proteins and surface curvature leading to a variety of membrane shapes dynamics \cite{fovsnarivc2019theoretical,gov2018guided}.
Here, we analyze this geometric influence on the stability region and patterning of the mechano-chemical system by using non-spherical initial geometries.
We chose to prescribe the cell shapes by Cassini ovals, the  geometry of which is given in Cartesian coordinates by the set of points $(x,y)$ which fulfill
\begin{align}
    \left( (x - a)^2 + y^2 \right) \left( (x + a)^2 + y^2 \right) &= b^4.
\end{align}
The parameters $a$ and $b$ control size and shape, varying the ratio $a/b$ allows to continuously tune the shape from a sphere to a constricted cell shape, with ellipsoidal-like shapes in between. 
%This describes the Cassini ovals with the two fixed points $(\pm a,0)$ and for each point on the curve the product of the two distances to the fixed points is equal to $b^2$ \todo{cite wolfram.com? better reference?}. 
In our case, we vary this ratio $a/b \in [0.2, 0.4, 0.6, 0.8, 0.95]$. The cell shape is defined as the rotational extrusion of this 2D oval, the size of which is scaled such that the enclosed volume is as before, i.e., equal to the volume of a sphere with radius R.  

The Cassini oval for $a/b\leq 0.6$ are similar to ellipsoids with increasing aspect ratios for increasing $a/b$. 
For $a/b = 0.8$ and $a/b = 0.95$ the shape becomes concave, alike a constricting cell. 
In numerical simulations with a freely deforming surface, we find that all Cassini oval shapes relaxes quickly towards a sphere. This can be explained as the initial isotropic surface contraction leads to increased concentration in regions of locally high surface curvature (consistent with the findings in \cite{Mietke2019-ju}). In consequence, the curvature-dependent surface tension $\xi f(c) H \mathbf{n}$ retracts the surface even faster than it would for a homogeneous concentration. 
Hence, to test the influence of cell shape on the pattern formation, we keep the geometry fixed in the following by using \cref{eqn:vn0}.

\begin{figure}
  	{\phantomsubcaption\label{fig:geometry_phasediagram}}%
  	{\phantomsubcaption\label{fig:geometry_shapes}}%
  	{\phantomsubcaption\label{fig:geometry_shapes_tubes}}
    \centering
    \includegraphics[width=0.8\textwidth]{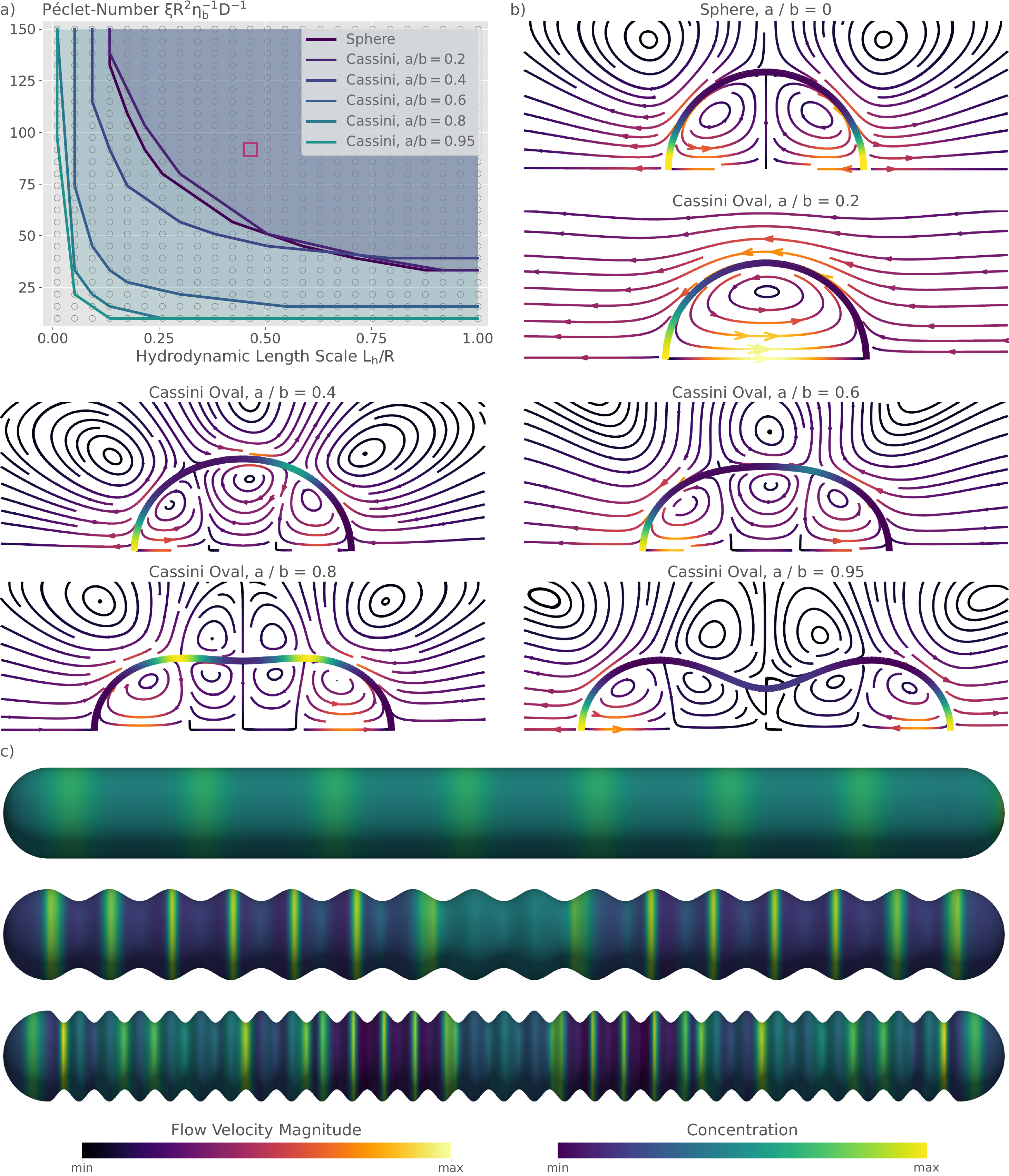}
    \caption{
    {\bf Geometrical Effect on Fixed Cassini Ovals}
    {\bf (a)}
    Phase diagram indicating the regions of instability (shaded areas) for the sphere and Cassini ovals $a/b \in [0.2, 0.4, 0.6, 0.8, 0.95]$. 
    The phase transition is approximated by the convex hull simulated points with pattern formation. 
    %The phase diagram for the fixed sphere is identical to the theoretical phase diagram as shown in \cref{fig:overview_phasediagram} up to the simulated resolution.
    {\bf (b)}
    Concentration patterns and flow profiles for $\pe = 91.7$ and $L_h = 0.46$ (purple marker in (a)) for the different geometries $\hat t = 10$ (stationary configuration).
    {\bf (c)}
    Patterning on elongated geometries:
    (top)
    plain elongated shape with a length of $16.5 R$ and a height of $0.75 R$,
    (middle)
    superimposed cosine wave with amplitude $0.1 R$ and period $2 \pi R$
    (bottom)
    same cosine wave but with period $4 \pi R$.
    The parameters for $\pe$ and $L_h$ are the same as in (b).
    The concentration is shown in log-scale. 
    Deviating parameters:
    $\omega = \num{1e-2}$,
    $dt / \tau_D = \num{5e-5}$,
    $h_\Gamma / R = \num{4e-2}$,
    in (c)
    $h_\Gamma / R = \num{2e-2}$   
    }
    \label{fig:geometry}
\end{figure}

\Cref{fig:geometry_phasediagram} shows the convex hull of the discrete unstable region for the fixed sphere and Cassini ovals with $a/b \in [0.2, 0.4, 0.6, 0.8, 0.95]$. 
The interface between stable and unstable region for the fixed sphere is the same as for the sphere with shape changes (compare to \cref{fig:overview_phasediagram}). 
For $a/b = 0.2$, the unstable region is slightly smaller compared to the sphere. 
For larger values of $a/b$, the unstable region increases in size. 
The concentration distributions and the flow profiles for each fixed shape is shown in \cref{fig:geometry_shapes} for the parameter set indicated by the purple marker in \cref{fig:geometry_phasediagram}.
Because our classification method is based on spherical shapes and the initial concentration distribution depends on the underlying discretization and thus geometry, we cannot make a statement about the resulting modes in general.
We, however, observed that the amount of higher order modes ($\ell^\ast\geq 3$) increases with increasing $a>0$. These modes are found to be even dominant at the phase transition (data not shown).
This would be consistent with the fact that the pattern formation has a preferable length scale, which increases the number of concentration peaks as the cell surface increases. 

To disentangle the influence of preferred length scale and geometry, we conduct simulations on tubular geometries with sinusoidal thickness profile (\cref{fig:geometry_shapes_tubes}). For a homogeneous tube (top) we find a clear preferred length scale (distance between rings). This length scale decreases significantly for the tubes with varying thickness as ring formations adapt to local curvature. Rings seem to preferentially develop in regions where the absolute value of local curvature is large (i.e., in regions of the highest positive or smallest negative curvature). 
A similar trend was previously reported for Turing patterns on curved surfaces, for which concentration spots emerge preferably in regions of higher local curvature \cite{Vandin2016-sv}.
The smaller distance between rings goes along with thinner ring structures and higher peak concentration.

We conclude that geometry and surface curvature have a strong influence on mechanochemical pattern formation and that investigations of active gel theory need to take the specific geometry into account to provide meaningful insights. 

% Conclusion & Outlook
\section{Conclusion}
We have presented a first numerical method for an active deformable surface interacting with the surrounding fluids. The underlying model couples surface and bulk hydrodynamics to surface flow of a diffusible species, which generates active contractile forces. We introduced a numerical discretization with moving finite-element grids and adapted it to the challenges of the problem: The pressure discontinuity across the surface is accurately resolved and surface mass and grid spacing are conserved under the occurring strong tangential flows. The introduced relaxation scheme allows a stable coupling between viscous stresses on the surface and in the surrounding bulk for a large range of hydrodynamic length scales. 
The method was validated with the linear stability analysis from \cite{Mietke2019-qf} and showed almost perfect agreement regarding predicted dominant patterning in the linear regime, near the phase transition. 

We further used the method to analyze the non-linear regime of spherical active surfaces. Spherical systems describe the generic shape of a surface under tension, like the cellular cortex \cite{Mayer2010-ex, Bray1988-nr, Berthoumieux2014-kb, Mokbel2020-wz}. Compared to the results from \cite{Mietke2019-qf}, we find higher order dominant modes.
The rich non-linear behavior of the system is also confirmed by the presence of multiple stationary states, which were found even with eliminated surface deformation.
Looking at the maxima in concentration and velocity of the obtained stationary states, we find that the mechanochemical instability is, in general, much weaker for symmetric patterns (even modes) with respect to asymmetric (polar) patterns. As a consequence, we did observe only slight constrictions at the formation of a contractile ring, far away from resembling cell division. This is also consistent with the previous finding, that Marangoni forces strongly suppress the constriction along a ring of increased surface tension \cite{De_Kinkelder2021-vp}. In more unstable parameter regimes (higher $\pe$ and $L_h$), contractile ring formations were observed only transiently, as they slipped away to the poles at later times. We conclude that an additional biochemical mechanism must be in place to induce the strong ring constriction observed during cell division. %Using an initial condition with increased concentration along the equator we observe such an event. 
Finally, we demonstrated that the method is capable to handle arbitrary closed surface geometries. 
We find that even small deviations from the spherical shape extend the mechanochemically unstable region significantly. On elongated geometries, ring-like patterns are more favored. Ring formations strongly adapt to local surface curvature, which largely changes the characteristic length scale of patterns. We conclude that geometry and surface curvature have such a strong influence on mechanochemical pattern formation, that studies of active gel theory need to take the specific surface geometry into account to provide meaningful insights.

The developed method provides a basis to analyze a variety of systems that involve mechanochemical pattern formation on active surfaces. Here, we focused on the simple case of a purely viscous fluid surface embedded in a viscous medium. 
{\color{black} Using full 3D simulations, we found that typically axisymmetric shapes and patterns emerge. %and restricted most parts of this study to axisymmetric scenarios. 
However, in the biological setting, this symmetry may be perturbed by intra- and extracellular cues or confinement. In the future, the method will be useful to investigate such non-axisymmetric processes}, such as cell migration or invagination events in tissues \cite{Bergert2015-qp, Martin2010-kl}.
Further, the constitutive relations of the surface can be readily adapted to account for different materials, for example the model can be combined with general models of viscoelastic surfaces \cite{De_Kinkelder2021-vp}. 
Also, an extension to more complex chemical scenarios would be interesting, for example involving several species which interact on the surface and in the bulk. Another interesting extension would be the inclusion of mechanosensitive binding dynamics. In recent years, several experimental studies have provided evidence that actin cross-linkers in the cell cortex exhibit such a dependency of attachment/detachment on mechanical cues \cite{Yao2013-pp, Schiffhauer2016-qf, Hosseini2020-cy}, which can be easily included in the model. Further, many studies suggested that active stress leads to shear-stiffening of biopolymeric networks \cite{Wang2002-ki, Koenderink2009-rz, Fischer-Friedrich2016-gs} due to cortical alignment and stress-induced structural changes. Extending the model to explore such non-linear effects and the associated complex pattern formation will be an important next step.

\textbf{Acknowledgements:}
We thank Marcel Mokbel, Simon Praetorius, Elisabeth Fischer-Friedrich and Mirco Bonati for support of the project and fruitful discussions on the topic. SA acknowledges financial support from the DFG in the context of the Forschergruppe FOR3013, project AL1705/6 and project AL1705/3. Simulations were performed at the Center for Information Services and High Performance
Computing (ZIH) at TU Dresden.

\appendix
\section{Appendix}

\subsection{Numerical Implementation}
\label{app:numerical_implemenation}

Here we describe some further details about the numerical implementation.
{\color{black}
We solve the \cref{eq:nondim1,eq:nondim2,eq:nondim3,eq:nondim4,eq:nondim5,eq:nondim6}
in a full three-dimensional scenario as well as in a rotationally symmetric scenario. 
Starting with the full three-dimension case, the computational domain~$\Omega$ is assumed to be a cylinder with length~$6R$ and radius~$3R$ containing a spherical surface with radius~$R$ in the center.
We assume Dirichlet boundary conditions $\mathbf{v} = \mathbf{0}$ at the cylinder surfaces and the pressure is fixed by $p = 0$ at the bases of the cylinder. 
No additional boundary conditions are needed for the concentration equation, as $\Gamma$ is a closed surface.
The weak problem given in \cref{eq:conc_weakform} can be implemented straightforward. 
However, to solve the bulk hydrodynamic weak problem with the explicit surface hydrodynamic term given in \cref{eq:hydro_weakform}, the surface forces need to be derived: 
For the surface tension term in Eq.~\eqref{bc stress discrete} we use that 
\begin{align}
\nabla_\Gamma\cdot \left(\left(\tilde{\gamma} + f(c)\right)P_\Gamma\right) =& 
\nabla_\Gamma \left(\tilde{\gamma} + f(c)\right) +  \left(\tilde{\gamma} + f(c)\right)\nabla_\Gamma \cdot P_\Gamma  \\
=&\nabla_\Gamma \left(\tilde{\gamma} + f(c)\right) + \left(\tilde{\gamma} + f(c)\right)H{\bf n}
\end{align}
where $H$ is the total curvature of the surface and $H\mathbf{n}$ is the total curvature vector. 
The latter can be computed by 
\begin{align}
H\mathbf{n} &= \Delta_\Gamma \mathbf{x} &\text{on}~\Gamma,
\end{align}
where $\mathbf{x}$ is the position on the surface~$\Gamma$.
The total curvature vector degenerates for $H = 0$.
Therefore, we approximate the normal vector $\mathbf{n}$ needed for the grid velocity~$\mathbf{w}$ in \cref{eq:grid_velocity} in the degrees of freedom by averaging the element normals of the neighbouring faces. 
}

To make the computational method more efficient, we {\color{black} also} implement an adapted scheme for rotationally symmetric scenarios. These include the biologically most relevant shapes and patterns on cell surfaces, such as polarization and ring formation \cite{Mietke2019-qf}.
Therefore, the effective computational domain reduces to two dimensions, similar to  \cite{Mokbel2020-uc}.
Computations are done on a 2D rectangular slice of extent $6R\times 3R$ of the cylindrical domain in the $x$-$y$ plane, where we assume the $x$-axis as the rotational axis.
To complete the numerical model, we assume Dirichlet boundary conditions for velocity and pressure on this rectangular domain: $\mathbf{v}_y = 0$ on all outer boundaries, ${\bf v}_x=0$ on the top boundary, $p = 0$ on the left and right boundary.
The active surface is placed in the center of this domain. At its two boundary points, the concentration field requires the symmetry condition $\partial_y c=0$.

To account for the missing third dimension, axisymmetric operators are used.
The axisymmetric gradient, divergence, and Laplace operators in the 2D coordinate system are defined by \cite{Mokbel2018-hw}
\begin{align*}
    \nabla = (\partial_x, \partial_y), \quad
    \tilde{\nabla}\cdot\star = \frac{1}{y}\nabla\cdot(y~\star) % (\partial_x, \partial_y+1/y)\cdot, 
    \quad
    \tilde{\Delta}\star = \tilde{\nabla}\cdot \nabla\star = \frac{1}{y}\nabla\cdot(y\nabla\star) % \partial_{xx}+\partial_{yy}+1/y~\partial_y
\end{align*}
and similarly for the surface operators
\begin{align*}
    \tilde{\nabla}_\Gamma\cdot\star = \frac{1}{y}\nabla_\Gamma\cdot(y~\star) % (\partial_x, \partial_y+1/y)\cdot, 
    \quad
    \tilde{\Delta}_\Gamma\star = \tilde{\nabla}_\Gamma\cdot \nabla_\Gamma\star = \frac{1}{y}\nabla_\Gamma\cdot(y\nabla_\Gamma\star)
    %\tilde{\nabla}_\Gamma\cdot = \left(\nabla_\Gamma+\begin{pmatrix} 0\\1/y \end{pmatrix} \right)\cdot, \quad
    %\tilde{\Delta}_\Gamma = \tilde{\nabla}_\Gamma\cdot \nabla_\Gamma = \Delta_\Gamma+ \begin{pmatrix} 0\\1/y \end{pmatrix}\cdot \nabla_\Gamma 
\end{align*}
These operators replace the corresponding divergence and Laplace operators in Eq.~\eqref{concentration time discrete} to yield the axisymmetric concentration equation. 
The axisymmetric hydrodynamic equations are similarly defined in the 2D computational domain, with the velocity field now being reduced to two components ${\bf v}=(v_x, v_y)$, and all tensors being reduced to 2x2 tensors.
The axisymmetric divergence operator replace the corresponding operator, but additional terms enter to account for the missing tensor components in 2D. 
Accordingly, in Eq.~\eqref{bulk stress balance time discrete}, we replace $\nabla\cdot S_i^n$ by 
\begin{align*}
    \tilde{\nabla}\cdot S_i^n + \begin{pmatrix} 0\\p_i^n/y \end{pmatrix}- \frac{\eta_i R}{\eta_b} \begin{bmatrix} 0&0\\0&2/y^2 \end{bmatrix}\cdot {\bf v}^n,
\end{align*}
and Eq.~\eqref{surface stress time discrete} is replaced by
\begin{align}
F_{\Gamma,{\rm visc}}^{n} = (1-\omega) F_{\Gamma,{\rm visc}}^{n-1} &+ \omega \tilde{\nabla}_\Gamma \cdot \left[(1-\nu)\tilde{\nabla}_\Gamma\cdot {\bf v}^{n-1} P_\Gamma +2\nu D_\Gamma^{n-1}\right] \\
&- \omega\begin{pmatrix} 0\\1/y \end{pmatrix}
\left[(1-\nu)\tilde{\nabla}_\Gamma\cdot {\bf v}^{n-1} +\frac{2\nu}{y} v_y^{n-1}\right]
&\text{on}~\Gamma 
\end{align}

\noindent
{\color{black}
We follow \cite{Mokbel2020-uc} to compute the total curvature~$H$ and normal~$\mathbf{n}$ in the axisymmetric scenario. 
}
%and in Eq.~\eqref{bc stress discrete} we replace $\nabla_\Gamma\cdot S_\Gamma^n$ by 
%\begin{align*}
%    \tilde{\nabla}_\Gamma\cdot S_\Gamma^n 
%    -(1-\nu)\tilde{\nabla}_\Gamma \cdot {\bf v}^{n-1} \begin{pmatrix} 0\\1/y \end{pmatrix}
%    -\nu\begin{bmatrix} 0&0\\0&2/y^2 \end{bmatrix}\cdot{\bf v}^{n-1}
%\end{align*}

An axisymmetric conservative weak form is derived by multiplication of Eq.~\eqref{eq:conc_weakform} by $y$: 
Find $c^n\in C_h$ such that for all $q\in C_h$,
\begin{align*}
0= \int_\Gamma &y\frac{c^n-c_{\hat{x}}^{n-1}}{dt} q - y c^n({\bf v}^{n-1}-{\bf w}^{n-1})\cdot \nabla_\Gamma q \\
&+ y c^n q \tilde{\nabla}_\Gamma \cdot {\bf w}^{n-1} + y\nabla_\Gamma c^n \cdot \nabla_\Gamma q + y\tau_D k_{\rm off}(c^{n}-1)q ~d\textbf{x}.
\end{align*}

The final discrete numerical system is implemented in the finite-element toolbox AMDiS (see \cite{Vey2007-af, Witkowski2015-yz}). 
The numerical domain is discretized on a one-dimensional surface grid for the cell surface and a two-dimensional bulk grid. Both grids match at the surface by construction. Grid size is set to $h_\Gamma$ (see \Cref{tab:params}) at the surface and coarser afar.
% The explicit treatment between surface viscous stress and bulk momentum conservation imposes time step restrictions which are handled by underrelaxation. The relaxation factor (see Eq.~\eqref{surface stress time discrete}) is set to $\omega=???$.
A preceding time step study was conducted to ensure time stepping errors are negligibly small at the chosen time step size of $dt \in [\num{5e-5}, \num{1e-4}]\tau_D$ depending on occurring stress magnitude.

{\color{black}
\subsection{Numerical Convergence}
\label{app:numerical_convergence}
We additionally validate our axisymmetric model numerically by computing the experimental order of convergence in space and time.
To this end, three uniformly refined meshes $M_0$, $M_1$, and $M_2$ with the maximal mesh size on the interface $h_{i, \Gamma} / R \in \{\SI{8e-2}{}, \SI{4e-2}{},\SI{2e-2}{}\}$  and three time-step sizes $dt_0$, $dt_1$, and $dt_2$ with $dt_i / \tau_D \in \{\SI{1e-4}{}, \SI{5e-5}{},  \SI{2.5e-5}{}\}$ are used.

We use four different error measures for different aspects of the system, namely the contour mean distance~$E_C$, the difference in the maximal concentration peak~$E_c$ and the differences in the flow fields in $x$- and $y$-direction~$E_{u_x}$ and $E_{u_y}$. 
The first measure, $E_C$, indicates the error regarding shape changes, the second regarding the concentration field and the last two regarding the flow fields.
They are defined by
\begin{align}
    E^{h_i\rightarrow h_{i+1}}_{C} &:= \frac{1}{N_0} \sum^{N_0 - 1}_{j = 0} \lVert\mathbf{X}^{h_i}_{2^i \cdot j} - \mathbf{X}^{h_{i+1}}_{2^{i+1} \cdot j}\rVert,
    \label{eq:passive:error_meandistance} \\
    E^{h_i\rightarrow h_{i+1}}_{c} &:= \left|\max{c^{h_i}} - \max{c^{h_{i+1}}}\right|,\\
    E^{h_i\rightarrow h_{i+1}}_{u_k} &:= \sqrt{\int_{\Omega^{h_i}_1 \cap \Omega^{h_{i+1}}_1} \left|u^{h_i}_k - u^{h_{i+1}}_{k} \right|^2}, \label{eq:passive:error_flow}
\end{align}
where $N_0$ is the number of vertices on the contour of $M_0$, $i \in \{0, 1\}$, $\mathbf{X}_j$ is the position of the surface grid point with index~$j$ (consecutively numbered) on the surface, and $k \in \{x, y\}$ depends on the velocity component in x- or y-direction.
We consider the vertices present in the coarser mesh only, and skipping every second (or every forth) vertex in the finer meshes.
The mesh convergence study considers only the inner phase for the flow field, as the outer fluid viscosity is almost negligible. 
For the convergence in time, the whole computational domain is considered as the mesh stays the same.

The experimental order of convergence is then derived for all four error measures~$E_e$ by
\begin{align}
    EOC_e &:= \frac{\ln E^{h_0\rightarrow h_{1}}_{e} - \ln E^{h_1\rightarrow h_{2}}_{e}}{\ln h_1 - \ln h_2}.
\end{align}
Note, that $\ln h_0 - \ln h_1 = \ln h_1 - \ln h_2 =\ln 2$, as we equally refined the mesh.
 
The resulting errors as well as the experimental order of convergences are listed in \cref{tab:active:eoc}. 
The experimental order of convergence for the mesh refinement is almost 2 for the cell shape (contour mean distance) as well as for the velocity fields $u_x$ and $u_y$. 
For the difference of the maximum concentration, we only get an order of 1. 
In time, we approximately get the convergence order of 1 as expected from an Euler scheme. 

The $L^2$ norms for the flow fields on the finest mesh $M_2$ are $||u_x||_{L^2} = \SI{3.47e0}{}$ and $||u_y||_{L^2} = \SI{1.24e0}{}$, which gives a relative error of $\SI{9.12e-1}{\%}$ for each direction.  
In time, the same norms for  $dt_2 / \tau_D$ are $||u_x||_{L^2} = \SI{1.85E+01}{}$ and $||u_y||_{L^2} = \SI{2.66E+00}{}$, which gives a relative error of $\SI{5.78E-03}{\%}$ and $\SI{3.25E-03}{\%}$, respectively. 
Thus, the difference is below \SI{1}{\%} for each here considered aspect, but without analytical results to compare with, stronger statements are not possible. 

\begin{table}[t]
\centering
\begin{tabular}{l|lll|lll}
$E_{e}$ & $h_0 \rightarrow h_1$ & $h_1 \rightarrow h_2$ & $\mathbf{EOC}^h_e$ & $dt_0 \rightarrow dt_1$ & $dt_1 \rightarrow dt_2$ & $\mathbf{EOC}^{dt}_e$ \\
  \hline
  \hline
  $E_{C}$ & \SI{8.38E-03}{} & \SI{2.77E-03}{} & 1.6  & \SI{3.86e-06}{} & \SI{2.36e-06}{} & 0.71\\
  $E_{c}$ & \SI{4.60e-02}{} & \SI{2.31e-02}{} & 1.0  & \SI{1.29e-05}{} & \SI{6.68e-06}{} & 0.95\\
  $E_{u_x}$ & \SI{1.00E-01}{} & \SI{3.16E-02}{} & 1.67  & \SI{2.13E-03}{} & \SI{1.07E-03}{} & 1.0\\
  $E_{u_y}$ & \SI{4.04E-02}{} & \SI{1.13E-02}{} & 1.84  & \SI{1.73E-04}{} & \SI{8.66E-05}{} & 1.0\\ 
\end{tabular}
\caption{
    {\color{black}
    {\bf Experimental order of Convergence:}  
    {\bf (left)}
        Errors and $EOC^h_e$ for the four different error measures~$e$ for meshes $M_0$, $M_1$, and $M_2$. 
        The time step was set to $dt_2 / \tau_D = \SI{2.5e-5}{}$.
    {\bf (right)}
    Errors and $EOC^{dt}_e$ for the four different error measures~$e$ for time steps $dt_0 / \tau_D$, $dt_1 / \tau_D$, and $dt_2 / \tau_D$. 
    The mesh was $M_2$ with $h_{i, \Gamma} / R = \SI{2e-2}{}$.
    Parameters:
    $\pe = 50$,
    $L_h / R = 1$,
    $\tau_D k_\text{off} = 10$,
    $\nu = 1$,
    $\tilde{\gamma} = 0$,
    $\eta_0 / \eta_b = \SI{1e-2}{}$,
    $\omega = \SI{1e-1}{}$.
    }
    }
\label{tab:active:eoc}
\end{table}
}

\begin{figure}
    \centering
    \includegraphics[width=0.8\textwidth]{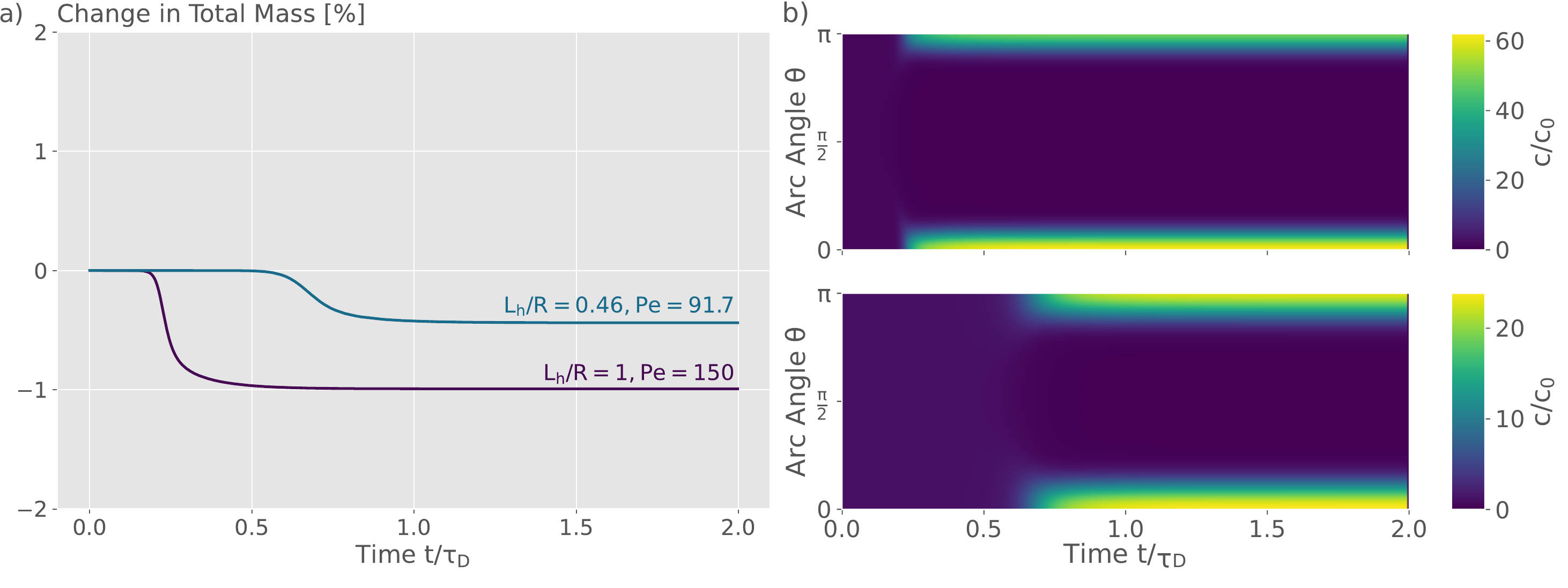}
    \caption{
    {\bf Numerical Validation}
    {\bf (a)}
    The relative change of the total mass $\int_\Gamma c$ either with mild ($L_h / R = 0.46$, $\pe = 91.7$) or strong ($L_h / R = 1$, $\pe = 150$) surface activity and dynamics under eliminated attachment/detachment,  $\tau_d k_\text{off} = 0$. 
    Error in mass conservation is below 1\% despite a complex pattern formation and shape dynamics (cf. \Cref{fig:evenmodes_slipping}) for a similar case).
    {\bf (b)}
    Kymograph of the two simulations in a). 
    Deviating parameters:
    $\tau_d k_\text{off} = 0$,
    $\omega = \num{1e-2}$,
    $dt / \tau_D = \num{5e-5}$,
    $h_\Gamma / R = \num{4e-2}$.
    %Simulation parameters: 
    %$\tau_d k_\text{off} = 0$,  $\nu = 1$,     $L_h = 1$,    $\pe = 150$,    $L^\prime_h = 0.01 \times L_h$,    $\tilde{\gamma} = 0$,    random seed = 1,    $dt = \num{1e-4}$,    $\omega = 0.01$,    $\max(h_\Gamma) = \num{1e-2}$.
    }        
    \label{fig:validation}
\end{figure}

Finally, we test the mass conserving property of the numerical implementation when attachment/detachment is switched off ($\tau_d k_\text{off} = 0$). \Cref{fig:validation} shows the mass loss for two test cases. For the strongest mechanochemical instability used in this paper ($\pe = 150$ and $L_h/R = 1$), we find a mass loss of about $1\%$. 
Note, that this is the accumulated value over the course of the full simulation. Once in the stationary state, mass is perfectly conserved despite the ongoing strong tangential flows. 
Note, that for most simulations presented in this paper we used $\tau_d k_\text{off}>0$, hence any local loss of mass is instantaneously balanced by attachment/detachment. 

\bibliographystyle{elsarticle-num}
\bibliography{main}

\end{document}